\documentclass[trackchanges,twocolumn]{aastex7}

\usepackage{colortbl}

\begin{document}
\begin{flushright} 
Accepted and scheduled for publication in ApJ
\end{flushright}
\title{Linear Polarization in a  Black Hole, 4U~1957+115 and  its  X-ray Spectral Analysis} 

\author[orcid=0000-0000-0000-0001,sname='Lev']{Lev Titarchuk}
\affiliation{INAF-IAPS Via Fosso Del Cavaliere 100, 00133, Rome, Italy}
\email{lev.titarchuk@inaf.it}


\author[orcid=0000-0001-5129-1491,gname=Seifina, sname='Elena']{Elena Seifina} 
\affiliation{Lomonosov Moscow State University/Sternberg Astronomical Institute,
Universitetsky Prospect 13, Moscow, 119992, Russia \email{seif@sai.msu.ru}}
\email{seif@sai.msu.ru}


\author[gname=Paolo,sname=Soffiitta]{Paolo Soffiitta}
\affiliation{INAF-IAPS Via Fosso Del Cavaliere 100, 00133, Rome, Italy}
\email{paolo.soffitta@inaf.it}

\author{Diana Makhinya}
\affiliation{Lomonosov Moscow State University, branch ``MSU Sarov'', Parkovaya st. 8, Sarov, Nizhniy Novgorod Region, Russia \email{dina185964@gmail.com}}
\email{dina185964@gmail.com}





\begin{abstract}

We present a new spectro-polarimetric method for estimation of the inclination of X-ray binaries using the linear polarization (LP) value in black hole (BH) sources, based on the early paper by Sunyaev \& Titarchuk (1985, ST85). The X-ray LP arises from multiple scattering of initially soft photons in a hot, optically thick Compton cloud (CC) using  a flat geometry. ST85 showed that the LP degree $P$ depends on the binary inclination 
$i$ and remains independent of the photon energy. Moreover, the $P$ follows a characteristic angular distribution based on the CC optical depth $\tau_0$, in particular for 
$\tau_0 >$ 10 the LP follows the Chandrasekhar (1950) distribution. In light of the vast number of recent IXPE spectro-polarimetric observations of BHs, testing this approach in combination with spectral analysis of these sources by other missions is desirable. In particular, the BH source, 4U~1957+115 was observed with IXPE in the high/soft state for which $P\sim 2\%$ was found (Marra et al. 2024). To apply the new method, 
in addition to knowing $P$ for a given source, it is necessary to estimate the CC optical depth $\tau_0$  and then to calculate of the best-fit photon index $\Gamma$ and the plasma temperature $kT_e$. We present the X-ray spectral analysis of 4U~1957+11 using data from {\it IXPE, NuSTAR, NICER, {\it RXTE}, Swift}, {\it Suzaku} and {\it ASCA}. We show that the X-ray source spectra are well described by the Comptonization model with $\Gamma$ varying from 1.5 to 3. We find a monotonic increase in $\Gamma$ with the mass accretion rate, $\dot M$, and a final saturation of the index  at $\Gamma\sim3$ for  high 
$\dot M$. We  determine a BH mass by the scaling method: $M_{1957}=4.8 \pm 1.8~M_{\odot}$, assuming a source distance of 22 kpc, using H~1743--322, 4U~1630--47, and GRS~1915+105 as reference sources.
\end{abstract}



\section{Introduction}

One of the important issues  of  astrophysics is an estimation of the masses of compact objects (COs) -- black holes (BHs), neutron stars, white dwarfs, since for them the mass is one of the most important available characteristics. In addition, to estimate the  CO mass 
 we need  to know an  orientation of such a binary  in space, that is, the inclination $i$ of the binary relative to an Earth observer. However, this  information is often missing due to the remoteness of the object  
 to estimate the parameters of a binary system \citep{Karttunen17}. It is known that the polarization of the radiation of the accretion disk surrounding the CO carries information about the inclination $i$ \citep[][hereafter ST85]{ST85}. 
In connection with the recent launch of the Imaging X-ray Polarimetry Explorer (IXPE) observatory, it became possible to measure the polarization degree of the radiation from 
black hole X-ray binaries (BHXRBs), which provides an alternative way to estimate the inclination $i$.

\begin{figure}
\includegraphics[width=8cm]{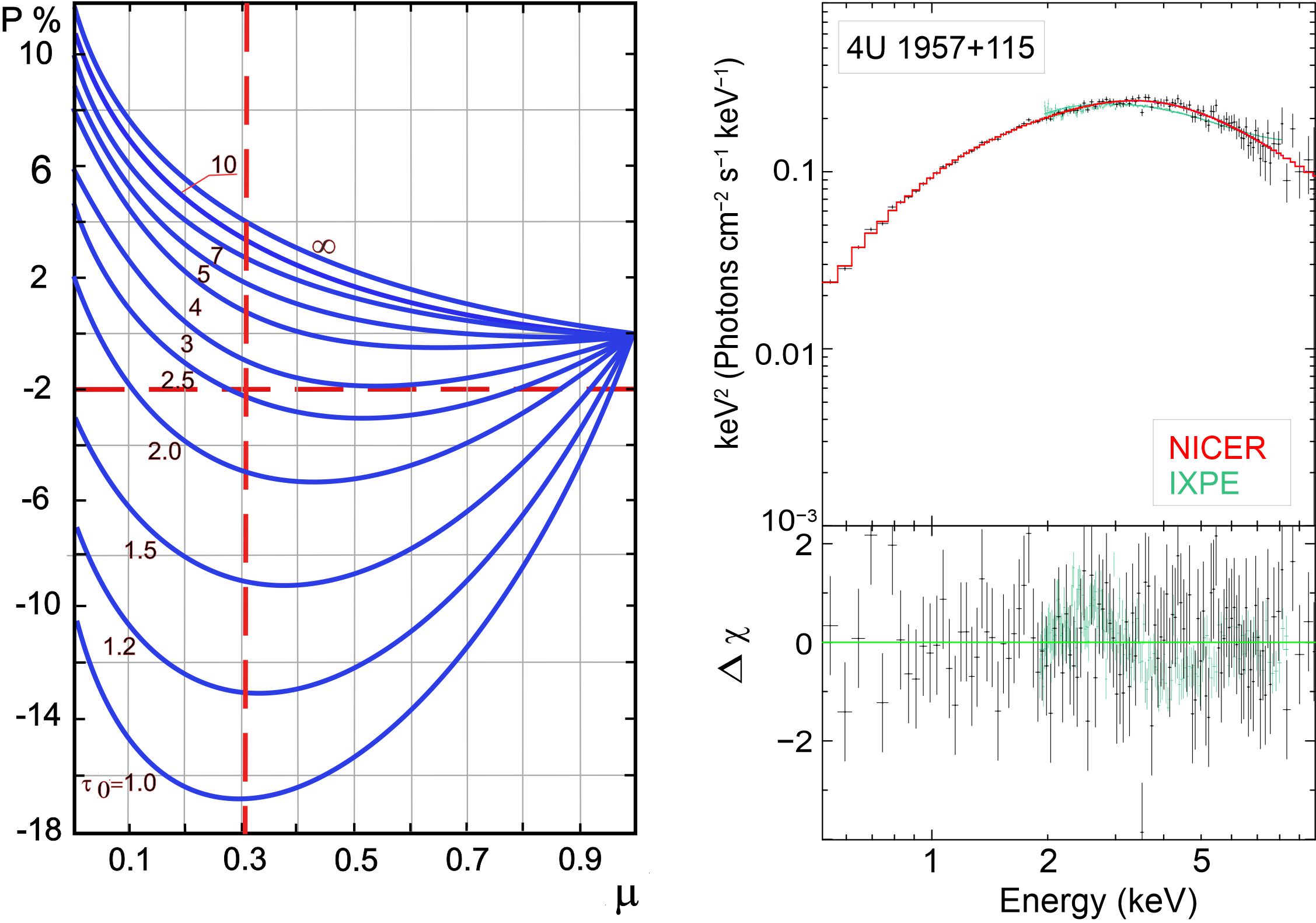}
\centering
\caption{Left: Hard-radiation polarization degree versus $\mu=\cos(i)$ for different $\tau_0$  (Figure is taken from ST85).  The values of $\tau_0$ are given next to the curves $P(\mu)$. Right: the X-ray spectrum of 4U~1957+115 observed by NICER (ID=610040101, red), which is simultaneous  one with the IXPE observation of 4U~1957+115  (ID=02006601, green). 
Best-fit parameters are  about
$\alpha=1$ (or the photon index $\Gamma=\alpha+1= 2$)  and $kT_e>10$ keV. 
}
\label{polariz}
\end{figure}

We  consider the problem of the X-ray spectral formation in the COs, in particularly  a BH, because of  exact measurement of the linear polarization (LP) in these objects.
In the presented  paper we  remind a reader ideas and  details of the  quite old  ST85 paper 
where the authors  studied, in details,   
the  LP value 
$P$ using the plane Compton cloud (CC) geometry for a wide range of the Thomson optical depth $\tau_0$  from 0.1 to more than 10. It is remarkable that  for $\tau_0$ 
(half of the slab) more  than 10  they obtained  a value of  $P$ which follows the classical Chandrasekhar distribution versus the inclination $i$ \citep{Chandrasekhar50} 
see,  for example, 
the case $\tau_0=\infty$ in Figure \ref{polariz}. It is well known that the form of observed X-ray spectra agrees  with  the analytically derived  Comptonization spectra \citep[see][hereafter ST80 and T94, correspondingly]{ST80,Titarchuk94}. 
Here, 
we interpret the X-ray data for  a 
BH in 4U~1957+115 and discuss the LP  established in this source [\cite{Kushwaha23},\cite{Marra24}, hereafter M24].

It is well known that BHXRBs undergo a change in different spectral states  \citep{McClintock+Remillard06}: from a low-luminosity hard state (LHS) through an intermediate state (IS) to a high-luminosity soft state (HSS).
These BHXRBs  demonstrate  the following  components in their spectra: a thermal component in the soft X-ray 
and a Comptonization component in the hard X-rays.  The thermal component is possibly formed  in  the accretion disk (AD), which is  interpreted as  geometrically thin disk of gas like  the Shakura-Sunyaev (SS)  disk   \citep[see in][]{SS73}. 
While the Comptonized component is formed   in the 
CC and  a hot CC plasma 
scatters photons from the accretion disk to higher energies, creating this  hard X-ray emission seen in  BHXRBs 
(see e.g. T94).

In the 
 HSS, the AD  black body dominates  in the emergent spectrum 
\citep[see e.g. the X-ray spectral evolution of  XTE~J1650--500 in][]{Montanari09}.   
The polarization data  give  an independent constraint on the X-ray emitting region. 
Therefore we applied  a new independent way  to check how accretion processes  can be  disclosed using   the polarization data. The IXPE  instrument 
\citep{Weisskopf22} allows us to measure  X-ray polarization  in  a wide range of stellar 
 BHXRBs  \citep[see e.g.][]{Saade24}.

\begin{figure*}
  \includegraphics[width=18cm]{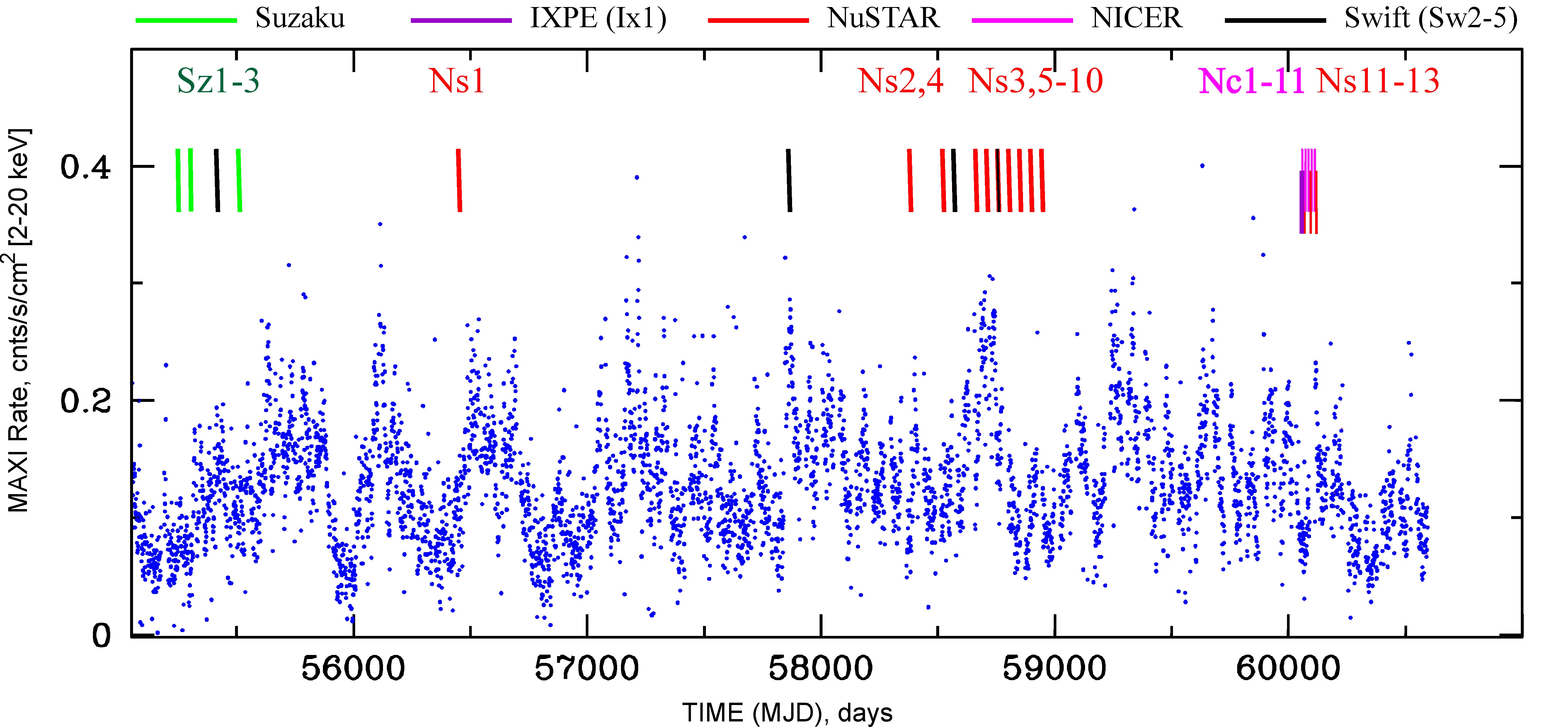}
\caption{Evolution of X-ray from 4U~1957+115 during  2009--2024 observations with {\it MAXI} (2--20 keV). Red vertical  lines (at top of the panel) indicate temporal distribution of the {\it NuSTAR} observations used in our analysis, whereas {\it Suzaku},  {\it NICER}, {\it IXPE}, and {\it Swift} observations, listed in Table~\ref{tab:table_Suzaku+ASCA_SAX}, are indicated by green, pink, purple, and black lines.
}
\label{ev_1957}
\end{figure*}

BHXRBs are excellent  laboratories for testing astrophysical models of X-ray spectral  formation.
 A particular interest is  a BHXRB with a  high luminosity in the soft X-ray range. 
 It is often difficult to diagnose the nature and parameters of the components of the binary system, since classical methods based on the orbital and/or eclipsing variability of the source, such as the dynamical method \citep{Karttunen17} and the eclipse method \citep{Tsesevich73}, are not applicable. A striking example of such a BHXRB is the source 4U~1957+115, discovered back in 1973 \citep{Giacconi74}. Five years later, \cite{Margon78} found its optical counterpart V~1408 Aql, which was identified as a binary system with an orbital period of $9.329\pm0.011$ h \citep{Thorstensen87,Bayless11}. One component of the system is apparently an evolved K2 or G2V secondary star \citep{Hakala14}, while the nature of the second component, which is assumed to be a X-ray source, is still unclear. Diagnostics of the nature of this second component (presumably, CO)  based on its mass, associated with measuring the mass function of 4U~1957+115, is very difficult because the object never enters a quiescent state (see an example of a typical X-ray light curve for 4U~1957+115 in Fig.~\ref{ev_1957}). Type I bursts, which would immediately indicate a neutron star (NS) as a CO, were not registered in 4U~1957+115.

%
%

%
%

%
%

\begin{figure}
\centering
 \includegraphics[width=8.5cm]{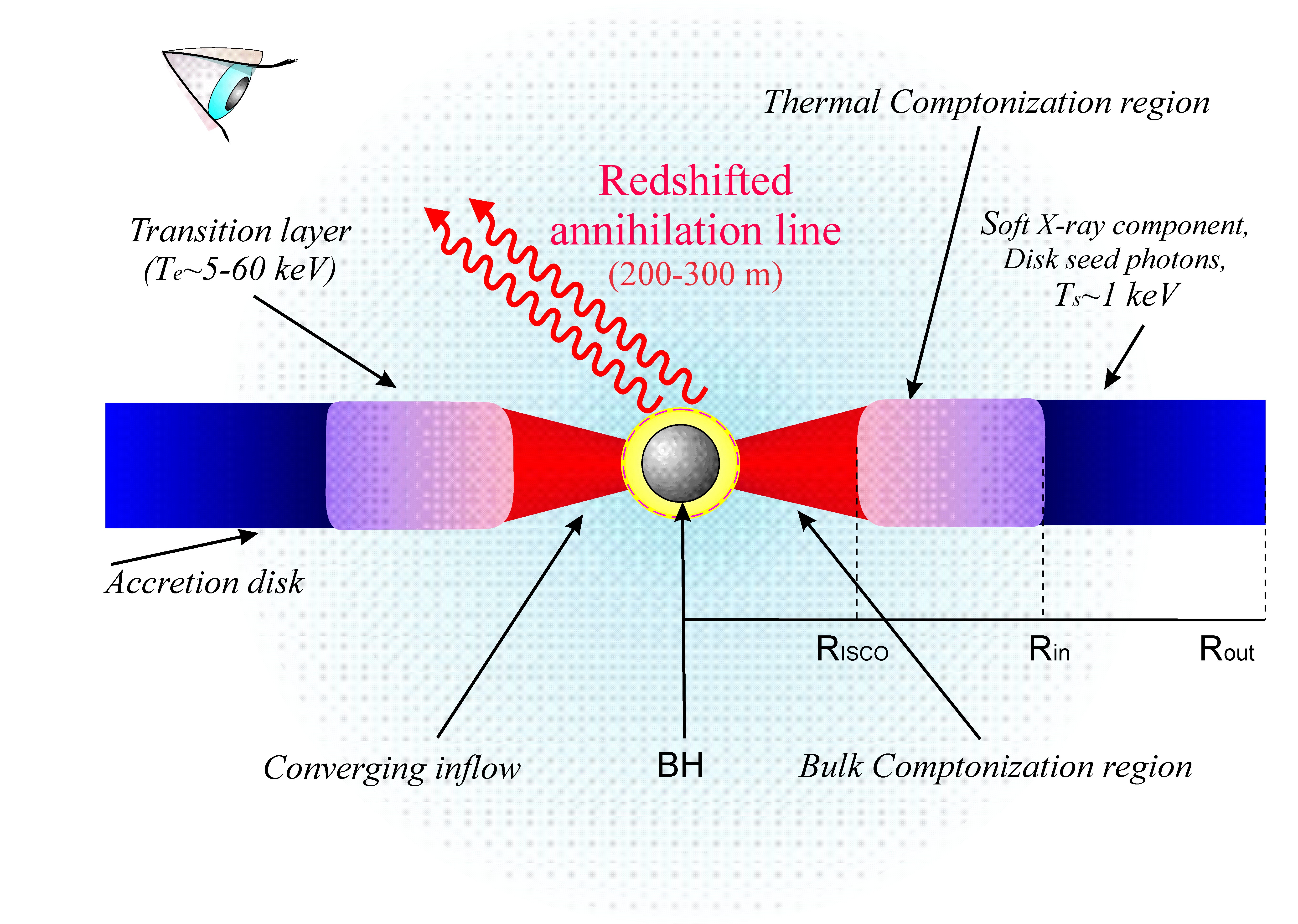}
\caption{A suggested geometry for 4U~1957$+$115. Disk soft photons are up-scattered (Comptonized) off relatively hot plasma of the transition layer. In addition, we consider the region near the BH horizon where photon-photon interactions lead to the pair production effect. In this process, the created positrons interact with the accreting electrons there, and therefore annihilation line photons are created and distributed over a relatively narrow shell (marked in yellow and shown enlarged in this plot) near the BH horizon. An Earth observer  should see this annihilation line only at gravitational redshift energies of $z \gg 1$.
}
\label{model}
\end{figure}

%
%


Therefore, other indirect methods for diagnosing the nature of the CO in 4U~1957+115 were used. Thus, \cite{Russell10} used the relationship between the optical and X-ray radiation of the source and classified 4U~1957+115 as a low-mass BHXRB. 
Based on the color diagrams, the source 4U~1957+115 was classified as a 
BH candidate in its soft spectral state \citep{Schultz89,White+Marshall84}  with dominance of the accretion disk radiation and very weak X-ray variability, 
which has persisted over the past 48 years \citep{Wijnands02}. The long-term X-ray flux variations in 4U~1957+115 with a period of 117 days was interpreted by \cite{Nowak+Wilms99}  as a precessional period of an edge-on accretion disk around a compact object \citep[BH or NS, see also][]{Yaqoob93,Wijnands02}. 
Another interpretation of this long-term X-ray variability is related to variations in the mass accretion rate through the disk due to instability caused by viscosity when the disk is irradiated by X-rays from its inner region \citep{Wijnands02,Ogilvie+Dubus01}. 

It is interesting that modeling of X-ray emission of 4U~1957+115 applied different models leads to significantly different results in the interpretation of the nature of this source. For an example  \cite{Singh94} and \cite{Ricci95} used the Comptonization model 
to describe the X-ray continuum using  EXOSAT data and concluded that there is a BH in 4U~1957+115. \cite{Singh94} traced an evolution of the Comptonization parameters and indicated that the plasma temperature, $T_e$ varies significantly from 12.5 keV (high-intensity state) to 1.3 keV (low-intensity one), although the optical depth  
varies from 4.5 (high-intensity state) to 22 (low-intensity state). At the same time, \cite{Singh94} did not exclude a presence of a weakly magnetized  NS. 
\cite{Yaqoob93} applied an additive 
model (consisting of {\tt bbody} with  a disk temperature $T_{in}\sim1.5$ keV and power-law components) to analyze the spectra of 4U~1957+115 obtained during {\it GINGA} observations, which also indicated that 4U~1957+115 was  a low-mass BHXRB. 
It is worth noting, however  that all these results strongly depend on an assumed distance to the source $d$ and an inclination $i$ of the object, which are strictly speaking not known. According to various estimates, $d$ varies from 7 kpc \citep{Sharma21} to 20--40 kpc \citep{Nowak08,Gomez15}. Monte Carlo simulations of simultaneous optical and near-IR photometry of 4U~1957+115 allowed  to make constraints on the distance $d$ to be imposed from 14 kpc to 80 kpc \citep{Hakala14}.

Some researchers reported extreme values of the inclination $i=78^{\circ}$ \citep{Maitra14} assuming a distance of 5--10 kpc based on simulations of Swift/XRT observations. On the other hand, \cite{Gomez15} reported an inclination angle $i$ of 13$^{\circ}$ 
in a model of the source consisting of a BH with  a mass of 3 $M_{\odot}$ and a normal  star companion with a mass of 1 $M_{\odot}$, assuming orbital modulation caused by variable irradiation of the companion star's surface during its orbital motion \citep{Thorstensen87,Bayless11}. Spectral analysis of {\it Chandra, RXTE}, and {\it XMM-Newton} observations of 4U~1957+115 within the relativistic accretion disk model showed that this source contained either a 3 $M_{\odot}$ BH at a distance of 10 kpc or a 16 $M_{\odot}$ BH at a distance of 22 kpc \citep{Nowak08,Nowak12}.

Therefore, additional criteria and new observations are needed to shed light on the nature of 4U~1957+115. Analysis of {\it RXTE} and ASCA observations of 4U~1957+115 by \cite{Nowak+Wilms99} pointed out   a NS, instead of  a BH.  However, the detection of a hard tail and weak variability during high luminosity states \citep{Bayless11,Wijnands02,Yaqoob93} can  speak in favor of both a BH and a NS. \cite{Barillier23} modeled the emission of 4U~1957+115 using the model {\tt eqpair} \citep{Coppi92} or the  Comptonization model {\tt thcomp} \citep{Zdziarski20} in combination with {\tt polykerrbb} \citep{Parker19} based on {\it NuSTAR} and {\it NICER} data. They suggest a BH with  a mass of 4.6 $M_{\odot}$ at a source distance 7.8 kpc \citep{GaiaCollaborationBrown21,Gaia CollaborationPrusti16}. In addition, they found a hard tail that exhibits a bidirectional behavior in  normalization vs Comptonization  parameter, $Y$ diagram 
\citep[see Fig.~8 in][]{Barillier23}: two tracks, both showing a hard tail increasing with 
X-ray flux, but one of them has significantly stronger tails.

Regarding a peculiarity of the X-ray spectrum of 4U~1957+115, the iron emission line at 6.5 keV is only occasionally present, but is often completely absent \citep{Nowak08}. The spectrum is also subject to interstellar absorption, $N_H=(1-2)\times10^{21}cm^{-2}$ \citep{Nowak08}. 

%
%

Recently,  
M24 \citep[see also][]{Kushwaha23} 
detected polarization of X-ray emission from 4U~1957+115 using IXPE observations. Namely, the LP 
degree for the source was 1.9\% $\pm$ 0.6\% and polarization angle (PA) was --41$^{\circ}$.8 $\pm$ 7$^{\circ}$.9 in the energy range of 2--8 keV. Spectral analysis of X-ray emission of 4U~1957+115 based on simultaneous observation by NICER (0.3--12 keV) 
and IXPE (2--8 keV) showed that the source was on average in the IS--HSS phase with a characteristic spectrum shape with a power-law spectral index of $1.93\pm0.21$ (M24). However, the source emission was somewhat variable (Fig. 1 in M24). 
This variability was clearly traced, for example, by the spectral hardness coefficient $HR$ (Fig.~2 in M24). 
Namely, $HR\equiv\frac{4-12 keV}{0.3-4 keV}$ 
according to NICER data varied from 0.05 to 0.06, and $HR\equiv\frac{5-8 keV}{2-5 keV}$
varied from 0.03 to 0.05 in IXPE data. 
At the same time, IXPE observations, indicate that the 
PA in 4U~1957+115  is independent of energy. 
Here we used a new independent way of analyzing mass accretion processes using X-ray polarization data to take a closer look at 4U~1957+115 in this state.


In Section \ref{polarization method} 
we provide a reader details of the polarization formation of the hard X-ray  emission in the plane CC 
around a BH.  
 We  use  the polarization properties of  4U~1957+115 to derive the main  parameters of this BH.  Sections \ref{data} 
-- \ref{results} 
provide the details of our data analysis and present a description of the spectral models used for fitting these data and results; Section \ref{discussion of spectral analysis} 
discusses the main results of the paper. In Section \ref{summary} 
we present our conclusions.

\section{The polarization problem  for X-ray photons \label{polarization method}}

\subsection{The linear polarization of X-ray photons in the plane Compton Cloud  
\label{polarization method_1}}
We assume that the primary source of photons is distributed over the CC slab at any law,  
either in the central plane of the CC slab, or consists of uniformly distributed sources in the CC slab, or at the boundary of the CC.

As ST85 demonstrated the final result for photons,  which undergo  a number of  scatterings  $N>>\tau_0^2$ much  more than average for a given optical depth of the CC slab, $2\tau_0$ (see Fig. 1 in ST85) is independent of these types of the  primary photon distribution. 

%
%


\begin{figure}
\includegraphics[scale=0.48,angle=0]{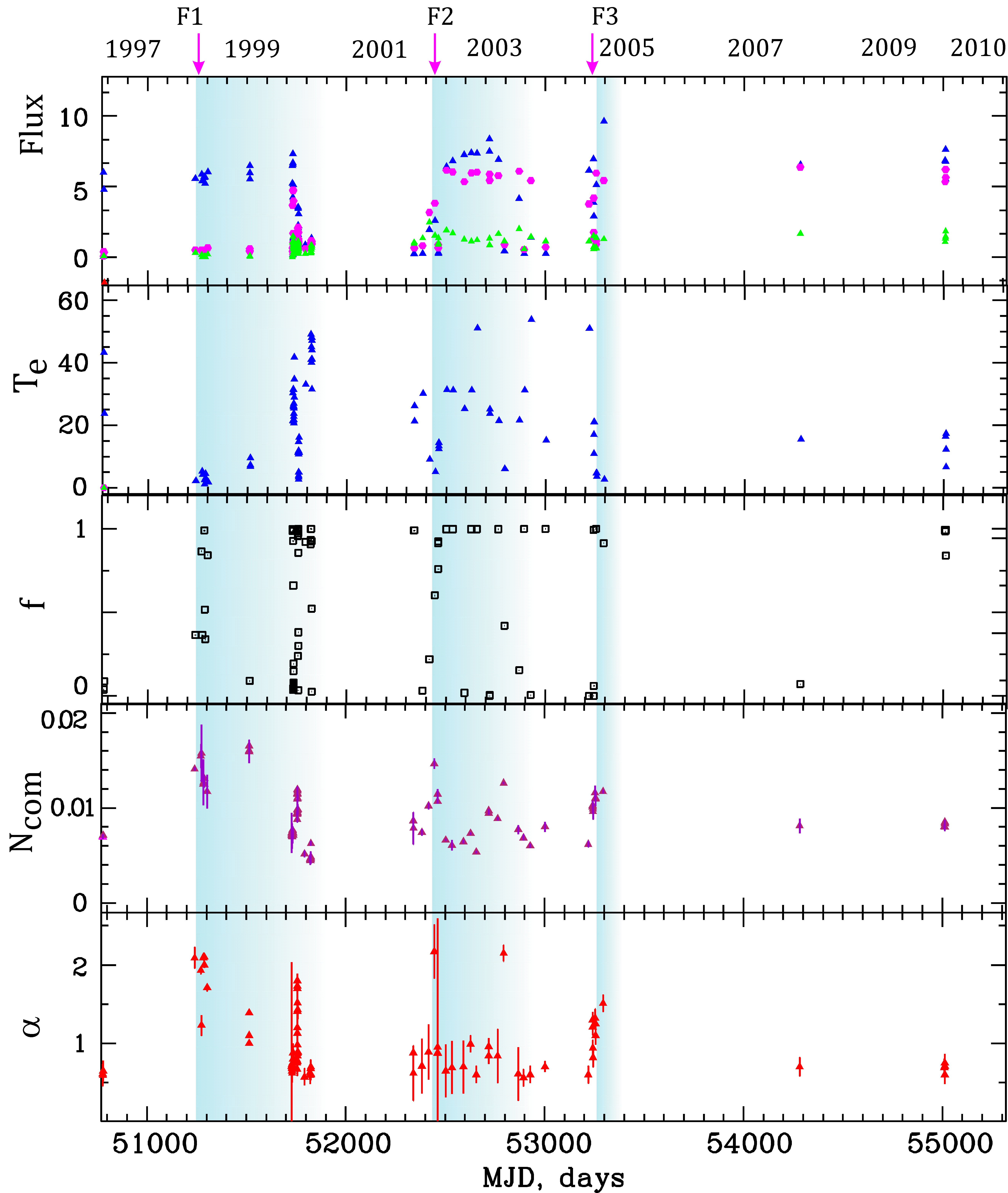}
\caption{
From top to bottom: evolution of the{ \it RXTE} flux  (in $10^{-12}$ erg/s/cm$^2$) 
in 3--10,  10--20 and 20--50 keV  ranges (blue, crimson and green points, respectively), 
electron temperature $T_e$ (in keV), Comptonized fraction $f$, and {\tt CompTB} normalization $N_{com}$  (in $L_{36}/d^2_{10}$) during 1997 -- 2010 flare events of 4U~1957+115. In the bottom panel, we present an evolution of the  spectral index $\alpha = \Gamma - 1$. The decay phases of the flares are marked with blue vertical strips. The peak outburst times are indicated by  arrows at the top of the plot. 
}
\label{fraq_1957}
\end{figure}

Upscattering (Comptonization) of low-frequency (soft) photons of  $h\nu\ll kT_e$ leads to energy gain up to photon energies, $h\nu\le 3 kT_e$.  It is a remarkable idea of ST85 
that the polarization of X-ray photons should be independent of the photon energy which is 
confirmed 
by the IXPE observations  (see the discussion on the constancy of $P(E)$ in observations below in Sect.~\ref{results_discussion_on polarization}). 


In order to solve  the integro$-$differential equation of the polarized radiation 
ST85 used the method of consecutive approximations. {\it For $k\gg\tau_0$ the polarization of photons after these $k-$scatterings  is independent of $k$}. When $kT_e< 70$ keV and  the Thomson approximation is valid the result of the ST85 calculations are correct \citep[see also ][]{Pozdnyakov83}.  
We denote $I_l$ and $I_r$ as the radiation intensities related to the electric field oscillations normal to the disk plane (and of direction of photon propagation) and in the plane perpendicular to it,  correspondingly.
Changes of  $I_l$ and $I_r$ in the CC plane  have been calculated  by the transport  equations [\cite{Chandrasekhar50}, see also, \cite{TSSCMM25}] 
\begin{eqnarray*}
\mu \frac{d}{d\tau} [I_l(\tau, \mu), I_r(\tau, \mu)]=[I_l(\tau, \mu), I_r(\tau, \mu)]\\
 - \int_0^1 P(\mu, \mu^\prime)
[I_l(\tau, \mu^\prime), I_r(\tau, \mu^\prime)] d\mu^\prime - [F_l(\tau), F_r(\tau)],
\end{eqnarray*}
\begin{equation}
\label{sc_matrix}
\end{equation}

where $d\tau=-\sigma_TN_e dz$, $\mu=\cos (i)$ and 

\[ P(\mu, \mu^{\prime})=\frac{3}{8}
\left\{
\begin{array}{ll} 
2(1-\mu^2)(1-{\mu^\prime}^2)+\mu^2{\mu^\prime}^2
 & \mu^2 
 \\
{\mu^\prime}^2 & 1
\end{array}
\right \}  \]
 is the scattering matrix. The vector ${\bar{F}} =(F_l, F_r)$ specifies  the distribution of primary sources. 

The boundary conditions of the problem are
\begin{equation}
{\bar{I}}(\tau=0, -\mu)= {\bar{I}}(\tau=2\tau_0, \mu)=0
\end{equation}
for any $0\leq \mu \leq1$.

The degree of the   polarization  is  defined as:
\begin{equation}
~~~~P=\frac{I_r-I_l}{I_r+I_l}
\label{pol_def}
\end{equation}
and the  observed intensity is a sum $I(0, \mu)=I_r(0,\mu)+I_l(0,\mu)$.

In order to find a 
 distribution for photons  which undergo $k$-scattering one should determine  the distribution of photons which are not scattered  in the medium. In this case one should solve 
 Eq.~(\ref{sc_matrix}) without the scattering term but with the initial  photon distribution $\bar{F}(\tau)$.

When the intensities of the first and next iterations  are calculated one can proceed with the intensity of photons which undergo $k$-scatterings:
\begin{eqnarray*}
\mu \frac{d}{d\tau} [I_l^{k}(\tau, \mu), I_r^{k}(\tau, \mu)]=[I_l^{k}(\tau, \mu), I_r^{k}(\tau, \mu)]\\
- \int_0^1 P(\mu, \mu^\prime)
[I_l^{k-1}(\tau, \mu^\prime), I_r^{k-1}(\tau, \mu^\prime)] d\mu^\prime.
\end{eqnarray*}
\begin{equation}
\label{k_iteration}
\end{equation}
The solution of the $k$-iteration depending of its solution on $(k-1)$ one (see also ST85).

In Table 2 of ST85 the authors showed that the polarization degree and angle distribution of the brightness of $k$-scattered  are independent of the primary source distribution.
To illustrate this statement the authors make their calculations for $\tau_0=2$ and  the iteration number $N\ge20$.
The calculation results for $k\gg1$ does not depend  on  $k$  and demonstrate that these results  can be applied  to the polarization of the Comptonized photons (see more details in ST85, in Appendix  A).  It is worth to emphasize that for large optical depths  ($\tau_0\gg 5$) X-ray polarization  converges to the  classical solution by  \cite{Sobolev49} and \cite{Chandrasekhar50}.

The appropriate radiative transfer equation (1) can be rewritten in the operator form (see also \S A2 in ST85).  The system of equations of the polarized radiation (1) can be rewritten as 
 \begin{equation}
{\bar{I}}=L{\bar{I}}+{\bar{F}}
\label{operator_form}
\end{equation}
where $L$ is the  polarization integral operator.
It can be checked that the operator $L$ meets all conditions  of the Hilbert-Schmidt 
 theorem [see e.g. \cite{Sobolev66}]. Thus, $L$ has a eigen$-$values ${\bar{p_i }}$ and 
ortho-normalized   eigen$-$functions ${\varphi}_i$ and $p_i\rightarrow0$ while $i\rightarrow\infty$.  A vector function ${\bar{F}}$  can be expanded in generalized Fourier series
 \begin{equation}
{\bar{F}}=\Sigma_{i=1}^{{\infty}} a_i {\bar{\varphi_i}}.
\label{vector_F}
\end{equation}

Solving Eq. (\ref{operator_form}) by the method of successive approximations we find the term ${\bar{I_k}}$ related to the photons  underwent $k-$scatterings:
 \begin{equation}
{\bar{I}}_k= \Sigma_{i=1}^{\infty} a_i p_i^k  {\bar{\varphi_i}}.
\label{I_k_term}
\end{equation}
Because of the sequence of $p_i$ decreasing with $k$  we obtain that 
 \begin{equation}
{\bar{I}}_k \simeq  a_1p_1^k{\bar{\varphi_1}}.
\label{final_form_I_k_term}
\end{equation}
This mathematical result is illustrated by Table 2 of ST85 for $k\gg \tau_0^2$.

\subsection{Determination of the main parameters of the Compton Cloud 
 \label{spectral analysis}}


The X-ray 
linear polarization 
$P$  is a result  of the multiple scattering  of the initially soft photons in the hot 
CC. Multiple scattering  of these soft photons leads not only to   the emergent  Comptonization spectrum  formation  but  also  these up-scattered X-ray photons become  linearly  polarized in the plane CC 
(see Fig. 1 in ST85).

A suggested geometry for 4U~1957$+$115 is presented in Fig. \ref{model}.  
We propose that accretion onto the BH occurs when the donor star material passes through two main zones: (i) a geometrically thin accretion disk (standard SS disk) and (ii) a transition layer (CC), where BH and disk soft photons are upscattered by hot electrons (Comptonized) by the relatively hot electrons of CC. The resulting thermal Comptonization spectrum is formed in the CC zone. 
In addition, we consider a zone near the BH horizon, where photon-photon interactions lead to the pair production effect. In this process, the created positrons interact with the accreting electrons there, and so annihilation line (AL) photons are created and distributed over a relatively narrow shell (marked in yellow and shown enlarged in this plot) near the BH horizon. An Earth observer can see this AL at energies much lower than those in the laboratory ($\sim$500 keV). Namely, only at energies taking into account the gravitational redshift $z \gg 1$ ($\sim$20 keV).

 In this consideration the polarization degree should be  independent of the photon energy for  photons which undergo  many scattering until they  reach  the 2$-$8 keV energy range of the  IXPE (see  \S 2 in 
ST80).
  
ST80 
(see also ST85) solved the problem of the X-ray spectral formation in the bounded medium. In particular, for the slab geometry they  derived a formula of the spectral index $\alpha$ (and $\Gamma=\alpha+1$):
  
 \begin{equation}
\alpha= -\frac{3}{2}
+\sqrt{\frac{9}{4}+\gamma},
 \label{sp_index}
 \end{equation}
 where $\gamma=m_ec^2\beta/kT_e$, 
$\beta=\pi^2/[12(\tau_0 +2/3)^2]$, 
and $\tau_0$ is  the optical depth of a half of the slab. 
 
 Using these formulas we can find the electron temperature  $kT_e$ as a function of $\alpha$ and $\beta$:
 \begin{equation}
 kT_e=\frac{\beta m_ec^2}{(\alpha+3/2)^2-9/4}.
 \label{kTe_alpha_beta}
 \end{equation}

ST85 calculated the  transfer of the polarized radiation using the iteration method. A number the iterations should be  much more than average number of scatterings in a given flat CC.  As a result they made  a plot of  the LP   as a function of $\mu=\cos (i)$, 
where $i$ is an inclination angle of the system for a different   Thomson optical  depths of  
 2$\tau_0$.    ST85 calculated   the LP  for $\tau_0$ from 0.1 to 10, and for $\tau_0>10$ they found that the LP follows the well known Chandrasekhar distribution \citep{Chandrasekhar50}. 
 
Let us recall that for a CO (BH or NS) in  4U 1957+115 
 $P=1.9\pm 0.6\%$ (M24), which is observed in the  IS--HSS. 
The source state was identified by modeling the spectrum with the best-fit parameter of the power-law index $\Gamma\sim 2$ (M24). Using the values of $P$ and $i\sim 70^\circ$ \citep{Maitra14}, we can estimate $\tau_0\sim 2.5$ (see Fig.~\ref{polariz}). 
Note that knowing the polarization, with a plot of LP against $\tau_0$, we can refine the inclination $i \sim 72^\circ$. 
 
 \subsection{Discussion of the polarization results}
 \label{results_discussion_on polarization}

It is known that observations  allow us to calculate the absolute value of polarization $|P|$ (or $|LP|$), but do not determine the sign of the polarization of the X-ray source. Therefore,
a  sign of the observed  polarization degree (see Fig.~\ref{polariz}), for example for  4U~1957+115, and the place of the LP in LP vs $\tau_0$ diagram  can be obtained using the following  arguments.
 
If we 
choose the  positive values, $P>2\%$  then we 
obtain that the inferred value, $\tau_0$ for $\mu=\cos (i) =\cos (70^\circ)\approx 0.3$. But this value of 
$\tau_0\sim 7 $  contradicts to the observed value of the index, $\Gamma\sim 2$ \citep{Marra24} or the  Comptonization parameter $Y\sim( kT_e /m_ec^2) (2 \tau_0)^2\propto \alpha^{-1}$. 
In fact, using  values of $\tau_0\sim7 $  and $kT_e>10$ keV (see Fig.~\ref{polariz}), 
 the emergent spectrum should have the Wien shape ($Y>1$), which contradicts to the  X-ray observations \citep{Marra24}. 
Namely, we should choose   a negative polarization value  of $P$ for 4U~1957+115.

As for the refinement of the CC geometry one can choose between   the spherical or the plane ones.  But  for the spherical one  $P$ is almost zero [compare  that  with  that in NS case \citep{farinelli24}]. It is worth noting that the  linear polarization can   be only   if there is a strong asymmetry   in the source. 
 The negative values of the polarization $P$ about 2\% for  4U~1957+115 
 leads  to estimates   of $\tau_0$  about 2.5 (see Fig. 5  in ST85 and Fig.~\ref{polariz} here).

Note also that  in practice, for many BHXRBs, the polarization degree $P$ is observed to increase with photon energy $E$ \citep[see e.g.][]{RodriguezCavero23,Ratheesh24,Krawczynski22,Kushwaha23a,Kushwaha23,Rawat23,Majumdar23a, Majumdar23b, Veledina23, Svoboda24a,Svoboda24b,Steiner24,Ingram24,Podgorny24,Majumdar25}. However, it should be noted that the increasing trend $P(E)$, for example, in 4U~1630--47 ranges from 6 to 12\% in the energy range 2--8 keV \citep[Table 1 in][]{Ratheesh24} and \citep{Kushwaha23}. On the other hand, the photon indices $\Gamma$ in these measurements range from 2.6 to 4.5 \citep{Ratheesh24,RodriguezCavero23}. This indicates that the polarization detection was made when the source underwent significant changes in the  spectral states. It is important to emphasize that the quasi-constancy of $P(E)$ 
can be  only  ensured if the $P(E)$ distribution is determined within the same spectral state. It is also not yet clear whether this change is caused by relativistic effects near the inner edge of the disk, while the polarization of the emerging disk radiation is energy independent. This remains to be confirmed with more observations, in fact, with more sensitive polarimetry towards 10 keV  and higher in future missions. 
But it is clear now that X-ray photons in order to be polarized they should be formed in the plane-like CC and undergoes quite a few scatterings, more than average. In this case the polarization properties of these X-ray photons depends on the geometry of the CC but not on the photon  energy.  

\cite{Saade24} (hereinafter S24) compare the X-ray polarimetric properties of stellar mass BHs and supermassive BHs. In their Fig.~1 they plot the polarization degree (PD, in $\%$) as a function of the BH inclination $i$. They found that the range of inclinations $i$ for Galactic BHs is well constrained (see above), while for extragalactic BHs (EBHs) they have a very broad distribution within about $60^\circ$. However, the PD values for these EBHs are typical of those calculated in ST85 and presented here in Fig.~\ref{polariz}. It is not by chance that S24 noted that the polarization properties for these two types of sources are similar. It is easy to see that the range of optical depths is $\tau_0\sim 1-1.5$.


 \section{Data selection \label{data}}
4U~1957+115 was   observed by {\it RXTE} (1996--2010), {\it Suzaku} (2010), $ASCA$ (1994), $Swift$  (2007--2019), {\it NuSTAR} (2013--2023), {\it IXPE} (2023) and {\it NICER} (2023). 
We extracted these data from the 
 HEAsoft (HEASARC 2014) 
archives and found that these data  cover a wide range of X-ray luminosities.  Data description see in \ref{suzaku data}--\ref{nustar data}. 
 4U~1957+11 was also observed with MAXI \citep{Matsuoka09}, whose light curve reproduces the quasi-continuous long-term variability of the source and helps to match pointed observations with the total flux in the 2--20 keV range (Fig.~\ref{ev_1957}).

\begin{table*}
 \centering 
 \caption{Details of ASCA,  {\it Suzaku}, {\it Swift},  {\it NuSTAR}, {\it IXPE} and {\it NICER} observations of 4U~1957+115. 
}
   \label{tab:table_Suzaku+ASCA_SAX}
 \begin{tabular}{llllcr}
 \hline\hline  
Epoch & Mission & Obs. ID& Start time (UT)  & Exposure (s) & MJD interval \\
      \hline
A1 & ASCA           & 42006000  & 1994 Oct 31 12:40:53 & 9867  &49656.5--49657.2$^1$ \\
 \hline
Sz1 & {\it Suzaku} & 405057010 & 2010 May 4 10:00:12 & 31158 
&55320.4--55321.4$^2$ \\
Sz2 & {\it Suzaku} & 405057020 & 2010 May 17 12:27:24 & 29509  
&55333.5--55334.4$^2$ \\
Sz3 & {\it Suzaku} & 405057030 & 2010 Nov 1 19:51:40 & 30317  
&55501.8--55501.9$^2$ \\
 \hline
Sw1 &	Swift &00030959001&2007 July 1 13:10:50	&182	&54282.5--4282.6$^5$\\
Sw2 &	Swift	&00091070001&2011 May 19 11:42:42	&61270&55700.4--55890.5$^5$\\
Sw3 &	Swift	&00030959006&2017 Apr 27 15:30:29	&925	&57870.6--57870.9\\
Sw4 &	Swift	&00088692001&2019 Mar 13 18:11:31	&5509	&58555.7--58556.6$^5$\\
Sw5 &	Swift	&00088975002&2019 May 9 01:23:27	&5448	&58731.--58818.1$^5$\\
\hline
Ns1 & {\it NuSTAR}  & 30001015002  & 2013 Nov 16 15:16:07 & 6379 &56612.6--56612.7$^{3,4,5}$ \\
Ns2 & {\it NuSTAR}  & 30402011002  & 2018 Sep 16 20:06:09 & 37250 &58377.3--58378.5$^{3,4,5}$ \\
Ns3 & {\it NuSTAR}  & 30502007002  & 2019 Apr 29 20:41:09 & 20730  &58602.8--58603.0$^{3,4,5}$ \\
Ns4 & {\it NuSTAR}  & 30402011004  & 2019 Mar 13 10:01:09 & 10110  &58555.4 --58555.8$^{3,4,5}$ \\
Ns5 & {\it NuSTAR}  & 30402011006  & 2019 May 15 12:01:09 & 37020  &58618.5--58619.8$^{3,4,5}$ \\
Ns6 & {\it NuSTAR}  & 30502007004  & 2019 June 4 19:56:09 & 20053  &58638.8--58639.1$^{3,4,5}$ \\
Ns7 & {\it NuSTAR}  & 30502007006  & 2019 July 19 06:11:09 & 18370  &58683.2--58683.4$^{3,4,5}$ \\
Ns8 & {\it NuSTAR}  & 30502007008  & 2019 Sep 10 01:36:09 & 10110  &58683.2--58683.3$^{3,4,5}$ \\
Ns9 & {\it NuSTAR}  & 30502007010  & 2019 Oct 20 12:51:09 & 19640  &58776.5--58776.7$^{3,4,5}$ \\
Ns10 & {\it NuSTAR}  & 30502007012  & 2019 Oct 30 20:56:09 & 20530  &58817.8--58818.0$^{3,4,5}$ \\
Ns11 & {\it NuSTAR}  & 30902042002  & 2023 May 15 08:01:09 & 18680  &60079.3--60079.5$^6$ \\
Ns12 & {\it NuSTAR}  & 30902042004  & 2023 May 19 00:21:09 & 20190  &60083.0--60083.2$^6$ \\
Ns13 & {\it NuSTAR}  & 30902042006  & 2023 May 24 05:41:09 & 20570  &60088.2--60088.4$^6$ \\
\hline
Ix1 &{\it IXPE}& 02006601  & 2023 May 12 02:41:23.184&  571490  &60076.1--60088.8$^7$\\
\hline
Nc1 & {\it NICER}  & 6100400101 & 2023 May 12 12:07:41& 657 &60076.54--60076.55 \\
Nc2 & {\it NICER}  & 6100400102 &2023 May 14 08:39:00 &2943&60078.36--60078.39 \\
Nc3 & {\it NICER}  & 6100400103 &2023 May 15 01:32:40&4820&60079.06--60079.12 \\
Nc4 & {\it NICER}  & 6100400104 &2023 May 16 02:19:40&8000&60080.09--60080.18 \\
Nc5 & {\it NICER}  & 6100400105 &2023 May 17 00:00:00&5072&60081.00--60081.06 \\
Nc7 & {\it NICER}  & 6100400107 &2023 May 19 00:29:40&4043&60083.02--60083.07 \\
Nc8 & {\it NICER}  & 6100400108 &2023 May 20 01:14:20&5651&60084.05--60084.12 \\
Nc9 & {\it NICER}  & 6100400109 &2023 May 20 00:14:40&8803&60085.01--60085.13 \\
Nc10 & {\it NICER}  & 6100400110 &2023 May 22 01:17:00&2926&60086.05--60086.09 \\
Nc11 & {\it NICER}  & 6100400111 &2023 May 23 00:22:20&4530&60087.01--60087.06 \\
      \hline
      \end{tabular}
\\Columns 1 -- 7 denote Epoch,  Mission Name, Observation ID, Start time (UT), Exposure time (s), MJD interval, respectively. See the text for details.
(1) \citet{Ricci95};  
(2) \cite{Nowak12}; 
(3) \citet{Barillier23};  
(4) \citet{Sharma21}; 
(5) \citet{Draghis23}; 
(6) this work; and 
(7) \citet{Marra24}; 
\end{table*}

%
%

\begin{table}
  \centering 
 \caption{The list of sets of  {\it RXTE} observation 
of 4U~1957+115.}
 \begin{tabular}{lllllc}
      \hline\hline
\\
Set  & Dates, MJD & Obs. ID&  Dates UT \\
                      &                  &                          &              \\
 \hline
\\
R1  &    50778--50781 & 20184$^{1,2}$                     & Nov 26, 1997 -- \\
      &                       &                                  &                      Nov 29, 1997   \\
R2  &    51240--51514 & 40044$^2$           & Mar 3, 1999 --  \\
      &                       &                       &                      Dec 1, 1999                        \\
R3  &    51729--52927 & 50128, 70014,            & July 4, 2000 --       \\
      &                       &     70054                      &                       Oct 15, 2003   \\
R4  &    53242--53294 & 90063, 90123$^3$,   & Aug 25, 2004 --   \\
      &                       &                        &                          Oct 16, 2004                    \\%
R5  &    54282.5--54282.7 & 93406$^2$           & July 1, 2007   \\
R6  &    55010--55013 & 94311$^2$   & June 28, 2009 --   \\
      &                       &                        &                         July 1, 2009                    \\%
R7  &    55321--55323 & 95335$^2$   & May 5, 2010 --   \\
      &                       &                        &                         May 7, 2010                    \\%
      \hline
      \end{tabular}
    \label{tab:par_RXTE_1636}
\\ References:
(1) \citet{Nowak+Wilms99}; 
(2) \citet{Wijnands02}  and   
(3) \citet{Nowak08}. 
\end{table}

%
%
\begin{figure*}
\centering
\includegraphics[scale=0.9,angle=0]{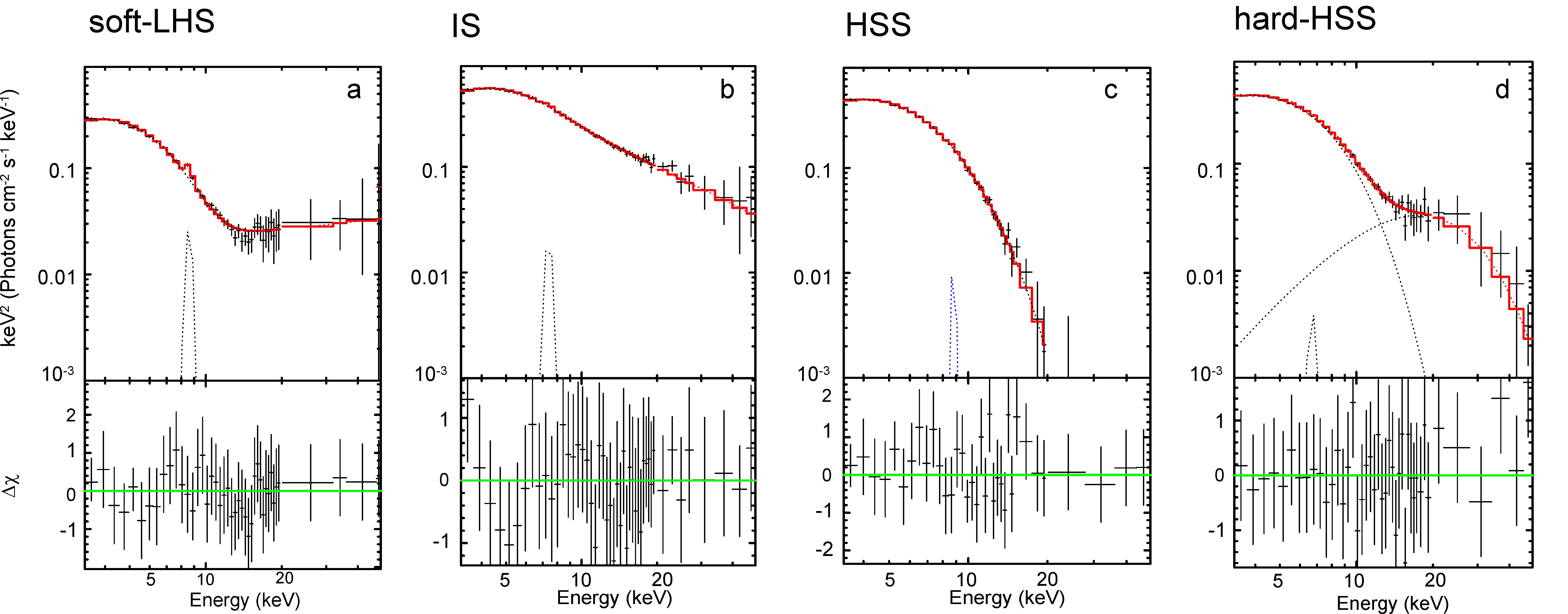}
\caption{Representative $E*F(E)$ spectral diagrams that are related to different spectral states of 4U~1957+115 using {\it RXTE} observation 50128-01-05-05 (soft-LHS), 70054-01-04-00 (IS),  40044-01-03-03 (HSS), and 70054-01-01-01 (hard-HSS)  with the best-fit modeling. The data are denoted by  crosses, while the spectral model is shown by a red histograms for each state.  Bottom: $\Delta \chi$ vs photon energy in keV. 
}
\label{spectrum_ev_1957}
\end{figure*}

%
%

\begin{table}
 \caption{Basic information on 4U~1957+115}
 \label{tab:parameters_binaries}
 \centering 
 \begin{tabular}{lllll}
 \hline\hline                        
Source parameter               &     Value \\
      \hline
Mass of primary star, M$_{\odot}$   & 3--20$^{~(a,b)}$ \\
Class of primary star & BH$^{~(b)}$, NS$^{~(c)}$ \\
Mass of secondary star  & 1$^{~(a)}$ \\
Class of secondary star & K2, G2V$^{~(d)}$ \\
Inclination,  $i$,  deg           & 13$^{~(a,d)}$--78$^{~(e)}$                         \\
Distance, $d$,  kpc       & 5--10$^{(e)}$, 10--22$^{(b)}$, 20--40$^{(d,e,f)}$       \\
                                  &  6--80$^{(g)}$          \\
RA (J2000), $\alpha$       & 19$^h$ 59$^m$ 24.1253$^{s~(h)}$      \\
Dec (J2000), $\delta$       & +11$^{\circ}$ 42{\tt '} 32.155{\tt ''}$^{~(h)}$        \\
Alternative names &  V1408 Aql,  \\
                          &2SXPS J195924.1+114231\\
                          & 1RXS J195924.2+114235 \\
$N_{H,Gal}$, cm$^{-2}$   &(1--2)$\times$10$^{21~(b,i)}$  \\
$m_V$, mag        & $\approx$19.0$^{~(d)}$  \\
Parallax $\pi$, mas & 0.07$\pm$0.15$^{~(j)}$  \\
 \hline                                             
 \end{tabular}
\\$^{(a)}$ \cite{Gomez15};
$^{(b)}$  \citep{Nowak08};
$^{(c)}$ \cite{Singh94}; 
$^{(d)}$ \cite{Hakala14};
$^{(e)}$ \cite{Maitra14};  
$^{(f)}$  \citep{Sharma21}; 
$^{(g)}$  \citep{GaiaCollaborationBrown21,Gaia CollaborationPrusti16}; https://www.swift.ac.uk/2SXPS/;
$^{(h)}$ \cite{Margon78}; 
$^{(i)}$ \cite{Marra24}; and 
$^{(j)}$ \cite{Kravtsov22}. 
\end{table}

\section{Results 
\label{results}}

%
%
\begin{figure}
\centering
\includegraphics[scale=1.24,angle=0]{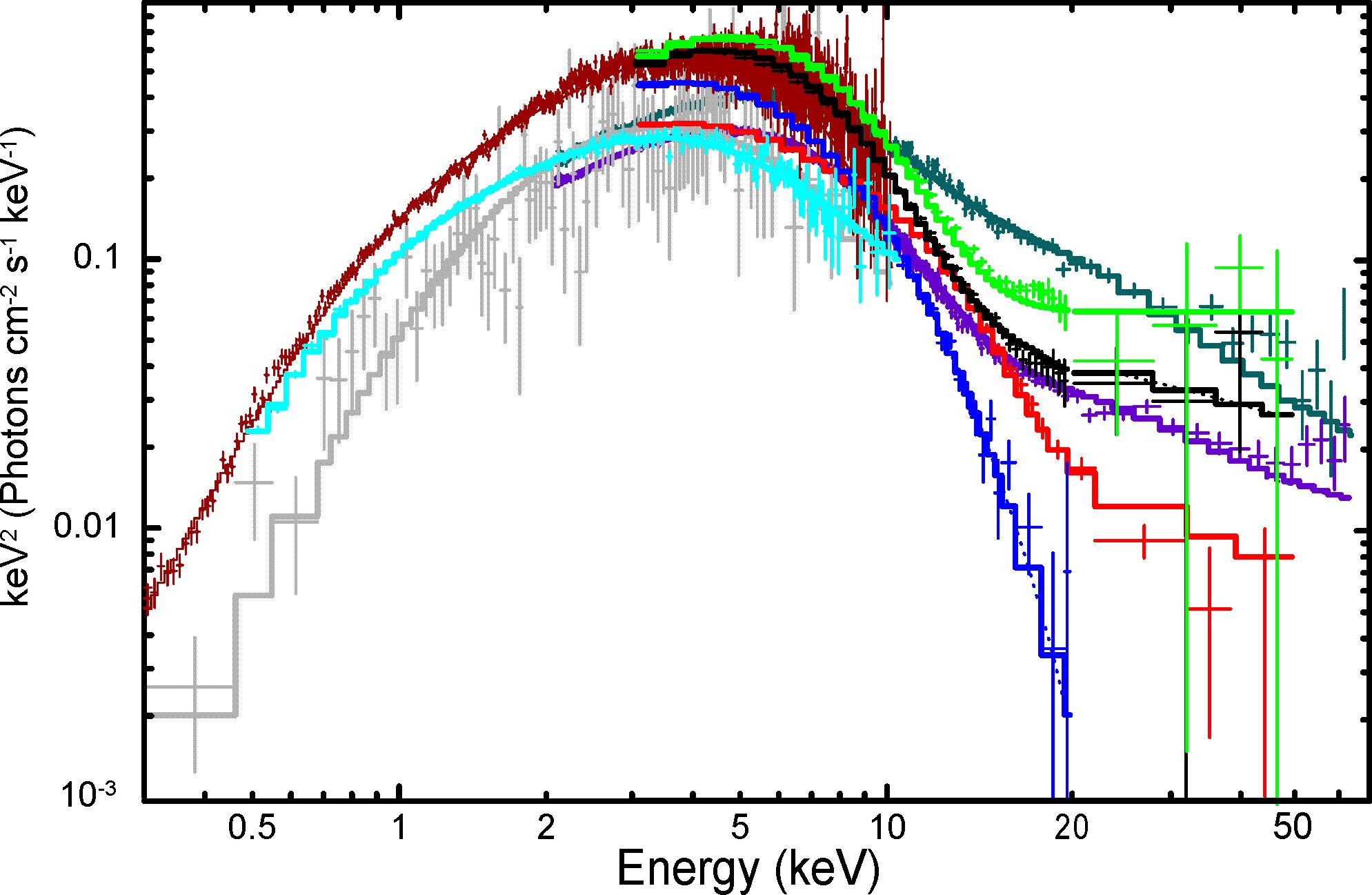}
\caption{
Representative spectra of 4U~1957+115 from {\it RXTE} data in units of $E*F(E)$ with the best-fit modelling for the LHS (ID=128-01-05-05, green), IS (ID=50128-01-05-02, black), hard-HSS (ID=50128-01-02-11, red) and HSS (ID=40044-01-03-03,blue) states. In addition, the {\it Swift} spectra of the source during the LHS  (ID=00030959001, gray),  and HSS (ID=00088975002, brown) states,  the {\it NICER} spectra during the IS  (ID=6100400101, bright blue) state, 
 as well as the {\it NuSTAR} spectra during  IS ($Ns8$, khaki) and HSS ($Ns5$, violet)
 are given.
%
}
\label{4_spectra_1957}
\end{figure}

%
%

\begin{figure*}
\centering
\includegraphics[scale=0.94,angle=0]{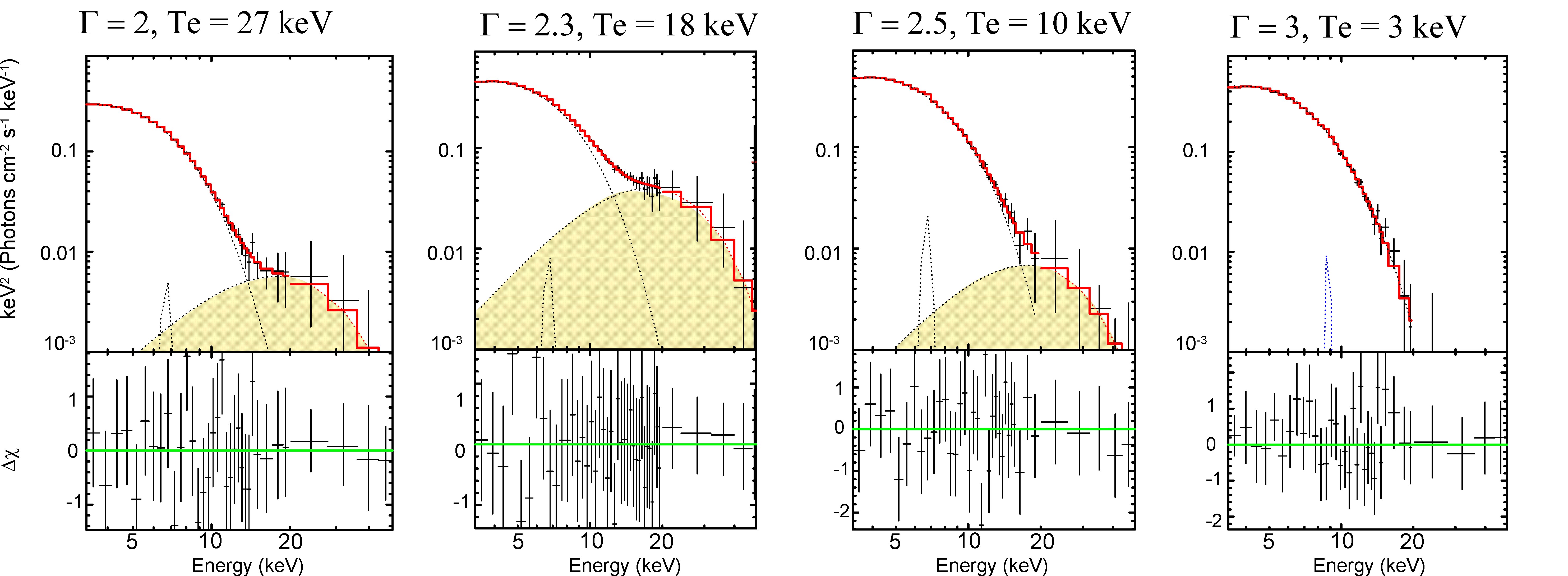}
\caption{Representative $E*F(E)$ spectral diagrams that are related to different spectral states for 4U~1957+115 using {\it RXTE} observations 50128-01-03-00 ($\Gamma=2\pm0.1$), 50128-01-02-11 ($\Gamma=2.3\pm0.1$),  70014-05-01-04 ($\Gamma=2.4\pm0.2$) and 
 40044-01-03-03 ($\Gamma=3.0\pm0.2$). The data are shown by black points and the spectral model is displayed by red histogram. Yellow shaded areas demonstrate an evolution of the HTBB component during spectral evolution between the IS and the HSS. 
}
\label{spectrum_ev_1957_HTBB}
\end{figure*}

\subsection{X-ray light curve\label{lc}}

To show the distribution of  {\it Suzaku},  $Swift$, {\it NuSTAR}, {\it IXPE} and {\it NICER} observations of 4U~1957+115, we have marked the times of their observations with green,   black, red, purple and pink vertical lines, respectively (see Fig. \ref{ev_1957}) along the MAXI light curve (2 -- 20 keV). The distribution of {\it RXTE} observations is shown in Fig.~\ref{fraq_1957}.


%
%

\begin{figure}
\centering
\includegraphics[scale=1.5,angle=0]{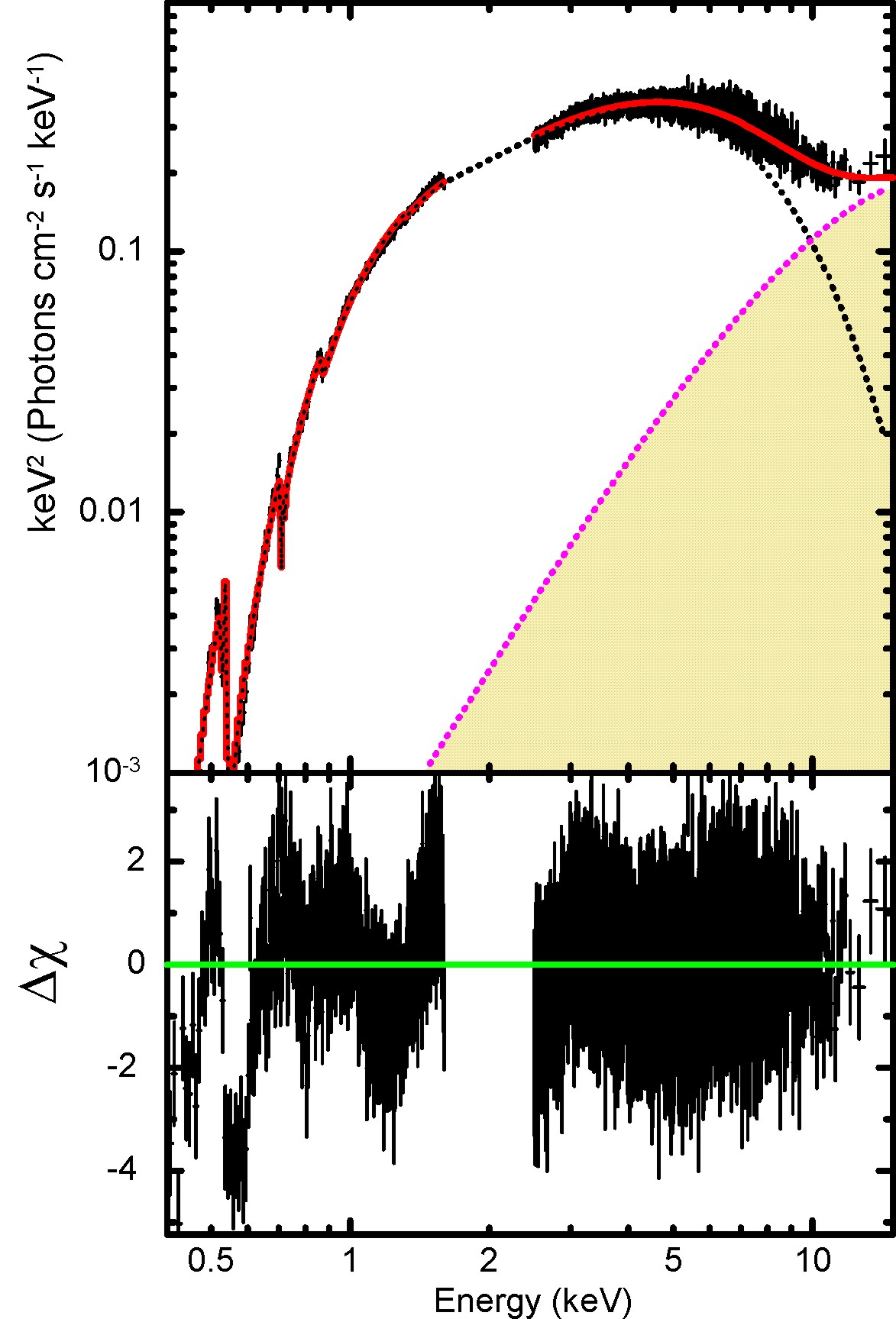}
\caption{$E*F(E)$ spectral diagram of 4U~1957+115 using {\it Suzaku} data in units of $E*F(E)$ with the best-fit modeling for the HSS (ID=405057010) state. The data are shown by black points and the spectral model is displayed by red hystogram. Yellow shaded areas indicates the HTBB component.
}
\label{Suzaku_spectra_1957}
\end{figure}


%
%

%
%
\begin{figure*}
\centering
\includegraphics[width=17.4cm]{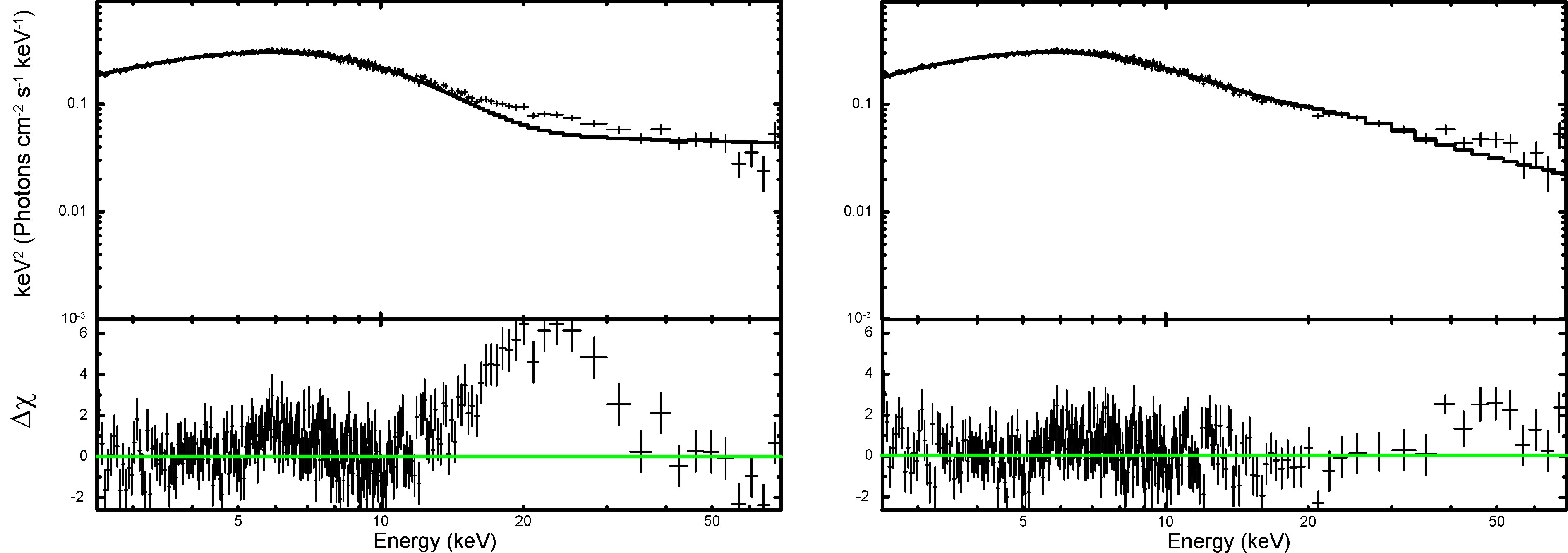}
\caption{The best-fit spectrum of 4U~1957+115 (top) with $\Delta\chi$ (bottom) during {\it NuSTAR} observation $Ns8$ in $E*F(E)$ units. Left: fitting without modeling the HTBB 
component ($\chi^2_{red} = 2.1$ for 552 dof) and right: the best-fit spectrum and $\Delta\chi$, when the bump in residuals at $\sim$ 20 keV is modeled by a HTBB component with $\chi^2_{red} = 1.08$ for 549 d.o.f. 
}
\label{NuSTAR-spectra_N8}
\end{figure*}

%
%
\begin{figure}
\centering
\includegraphics[scale=0.75,angle=0]{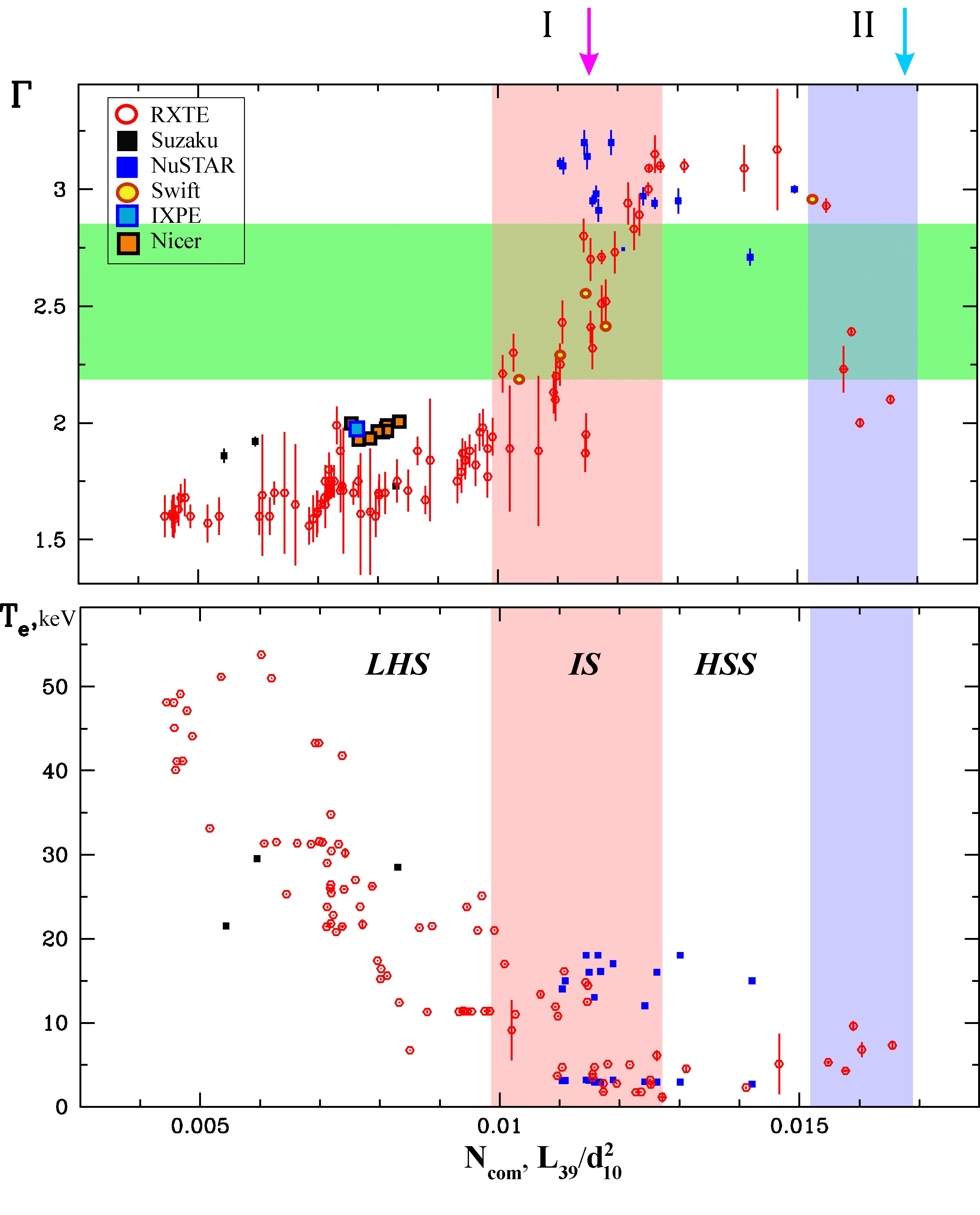}
\caption{The photon index, $\Gamma$ plotted versus the $CompTB$ normalization, $N_{com}$ (top panel) and  the electron temperature $T_e$ (in keV, bottom panel) for 4U~1957+115. Here, two zones (I and II) for $\Gamma\sim$ 2.2--2.8 are also highlighted, where the appearance of the HTBB is expected (green band for $\Gamma$). The photon index, $\Gamma$ decline at large {\it CompTB} normalizations is indicated by a vertical blue stripe (zone II). 
 The pink vertical band (zone I) marks the normalization interval corresponding to the detection of the HTBB at average accretion rates $\dot M$ ($\sim N_{com}$).
}
\label{gam_norm_Te}
\end{figure}


%
%
\begin{figure*}
\centering
\includegraphics[width=17cm]{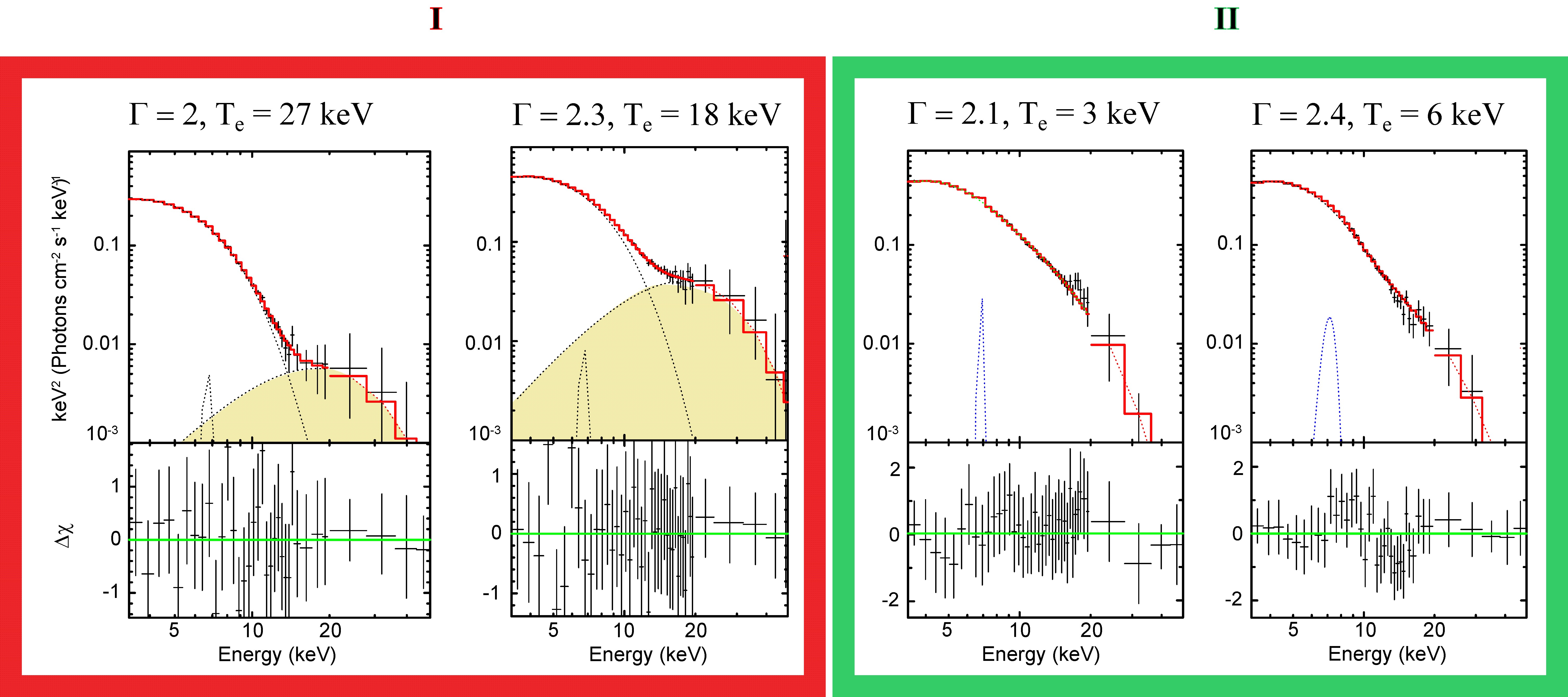}
\caption{Difference between the spectra of 4U~1957+115 in zone I (enclosed 
by a red rectangle) and zone II (enclosed 
by a green rectangle) of 
the $\Gamma-N_{com}$ diagram in Fig.~\ref{gam_norm_Te}. It is evident that in zone I the contribution of the HTBB is well detected, although in zone II the spectrum is well described by the Comptonized model without any contribution of the HTBB.
The best-fit spectra of 4U~1957+115 using {\it RXTE} observations (in zone I)  50128-01-03-00 (left) and 50128-01-02-11 (right); (in zone II) 40044-02-01-00 (left) and 40044-02-01-01 (right) are used to demonstrate this difference.  The data  are presented by crosses and the best-fit spectral  model   {\it tbabs*(CompTB+BB)} by red line. Yellow shaded areas demonstrate an evolution of the HTBB component during spectral evolution between the LHS and the HSS.  Bottom: $\Delta \chi$ vs photon energy in keV. 
}
\label{HTBB_I_II}
\end{figure*}

\subsection{Spectral analysis \label{spectral analysis}}


The broadband spectrum of 4U~1957+115 was simulated within the XSPEC v12.14.1 software package using an additive model consisting of the Comptonization component CompTB \citep{ft11} and the iron K$_{\alpha}$ emission line at 6--8 keV described by the Gaussian model. The CompTB component is applicable to any state and Comptonization type (thermal or dynamic) for the observed X-ray spectra of BHs or NSs \citep{TS24,TSC23,TS23,TS21,TSCO20,TS17,TS16,tsei16b,STS14,ft11,tss10,ts09} and 
describes the continuous X-ray spectrum of the source formed as a result of Comptonization of soft ``seed'' photons in a CC 
of hot electron plasma located near the compact object. 
In addition, the {\tt CompTB} model calculates the soft excess and the primordial emission self-consistently. In this model, see Figure {\ref{model}}, 
the total emission is determined by the {\tt CompTB} normalization $N_{com}$, which is proportional to a mass accretion rate $\dot M$ and the spectral index $\alpha$ (or the photon index $\Gamma=\alpha+1$). The disk emission appears as the  blackbody emission with a color  temperature $T_s$  at radii $R_{TL}<R<R_{out}  $, where $R_{out}$ and $R_{TL}$ are the outer radius of the AD 
and the outer radius of the transition layer (TL), respectively. At $R<R_{TL}$, 
the AD emission comes out as the Comptonized emission from the warm and optically thick medium, rather than as  the thermal one. The hot and optically thin TL is located in the inner part of the AD around a BH ($R <R_{TL}$), and this region forms  a high-energy power-law continuum. The total Comptonized radiation is a result of  the Comptonization in  the X-ray energetic TL and the  BH-converging inflow ({\it Bulk Comptonization region}), 
where thermal and dynamical  Comptonizations are effective).  A fraction of the   Comptonized radiation ($f=A/(1+A )$ or $\log A$) can be found using this  model fit. The seed photon temperature $T_s$ and the TL spectral index of the corona determine  the energy of the upward-scattered  off soft excess radiation. The plasma temperature of accretion flow  $T_{\mbox{e}}$ (temperature of the electrons in keV) is also the parameter of the {\it CompTB} model. The Comptonization continuum can be calculated  as a convolution of the soft Blackbody radiation with  the Comptonization Green function $G$, which can be approximated  as
\begin{eqnarray}
G(x,x_0) & = & x^{-\alpha} , ~~~\rm {for}~~x>x_0, \nonumber\\
G(x,x_0) & = &  x^{\alpha+3}, ~~~\rm {for}~~ x<x_0, \nonumber
\end{eqnarray}
\noindent where $x = hv/kT_e$, $x_0$ is the breakpoint of $G(x,x_0)$  where the two cases ($x>x_0$ and $x<x_0$) meet each other. Therefore, the general model consists of a Blackbody-like (BB) and a Comptonized component (XSPEC models ``BMC", 
``COMPTB", ``COMPTT" are the sum of these components: $BB+f\cdot BB*G$). 
The bulk parameter of {\it CompTB} model, $\delta$, which indicates  an efficiency of the bulk motion with respect  to the thermal Comptonization one, for simplicity we assume 
as $\delta=0$.


For 4U~1957+115  we used the Comptonization model {\tt CompTB} modified 
by 
the additive {\it Gaussian} line at $\sim$~6.5 keV.
The parameters of a  {\it Gaussian} component are  a centroid line energy $E_{line}$, the width of the line $\sigma_{line}$  and normalization, $N_{line}$ to fit the data in the 6 -- 8 keV  energy range.  We also apply  interstellar absorption ({\tt tbabs} model) with  column density $N_H$  (see Table~\ref{tab:parameters_binaries}). The resulting model reads in XSPEC as 
{\tt tbabs*(CompTB+Gaussian)}.


%
%
\begin{table*}
  \centering 
 \caption{Best-fit parameters of spectral analysis of some {\it ASCA}, {\it Suzaku}, {\it Swift}, {\it NuSTAR}  and {\it NICER} observations of 4U~1957+115$^{\dagger}$.
}
 \begin{tabular}{lccccccccrcccccc}
      \hline\hline
Epoch & $\alpha=$     & $T_{s}$,   & $T_e,$ & $\log(A)$ & N$_{com}^{\dagger\dagger}$ &$E_{gauss}$, &$N_{gauss}^{\dagger\dagger}$ &$N_{htbb}^{\dagger\dagger}$ &  $\chi^2_{red}$ \\
             &  $\Gamma-1$ & keV          & keV      &               &                                              &               keV                            &  & &         (dof) \\
\hline 
\rowcolor[cmyk]{0,0,1,0}A1&  0.73$\pm$0.04& 0.40$\pm$0.02&  18.52$\pm$0.02 & 2.00$^{\dagger\dagger\dagger}$ &6.69$\pm$0.03 & 7.01$\pm$0.06&0.01$\pm$0.01& 0.00001(1) & 1.07(277) \\
\hline 
Sz1&
 0.73$\pm$0.01& 0.561$\pm$0.005 & 28.52$\pm$0.02  &2.00$^{\dagger\dagger\dagger}$ &8.30$\pm$0.03  &6.64$\pm$0.08&0.02$\pm$0.01& 0.00001(1) & 1.03(2492)  \\
Sz2&
 0.92$\pm$0.02& 0.557$\pm$0.005&  29.52$\pm$0.02&  2.00$^{\dagger\dagger\dagger}$&5.95$\pm$0.03&6.78$\pm$0.09&0.01$\pm$0.01&  0.00001(1) & 1.12(2297) \\
Sz3&
 0.86$\pm$0.03& 0.561$\pm$0.006&  21.52$\pm$0.02&  2.00$^{\dagger\dagger\dagger}$ &5.43$\pm$0.03&  6.52$\pm$0.03&0.01$\pm$0.01&0.00001(1) & 0.96(2244) \\
\hline 
Sw1& 2.05$\pm$0.04& 1.84$\pm$0.09&  7$\pm$0.2&  2.00$^{\dagger\dagger\dagger}$ &15.08$\pm$0.01&  6.4$\pm$0.2&0.01$\pm$0.01&0.00001(1) & 1.21(334) \\%
Sw2& 1.3$\pm$0.1   & 1.12$\pm$0.09  &  10$\pm$0.1&  -0.14$\pm$0.02 &10.5$\pm$0.2                             &  6.5$\pm$0.1&0.01$\pm$0.01&0.00001(1) & 1.19(933) \\%
Sw3& 1.4$\pm$0.2   & 1.14$\pm$0.01  &  11$\pm$0.1&  -0.16$\pm$0.03 &11.1$\pm$0.2                             &  6.4$\pm$0.2&0.01$\pm$0.01&0.00001(1) & 1.23(927) \\%
Sw4& 1.3$\pm$0.2   & 1.15$\pm$0.01  &  12$\pm$0.2&  -0.12$\pm$0.01 &12.98$\pm$0.06                          &  6.4$\pm$0.1&0.01$\pm$0.01&0.00001(1) & 1.02(994) \\%
Sw5& 1.4$\pm$0.1   & 1.17$\pm$0.03  &  15$\pm$0.2&  -0.15$\pm$0.02 &11.9$\pm$0.1                            &  6.4$\pm$0.1&0.01$\pm$0.01&0.00001(1) & 1.07(994) \\%
\hline 
\rowcolor[gray]{.9}Ns1&
2.00$\pm$0.03&  1.01$\pm$0.01 & 18.6$\pm$0.1  &-0.76$\pm$0.01 &11.25$\pm$0.02  &6.95$\pm$0.03&0.02$\pm$0.01&0.0003(1) & 0.99(424)  \\
\rowcolor[gray]{.9}Ns2&
1.95$\pm$0.04&  0.98$\pm$0.02 & 13.1$\pm$0.2 &-0.77$\pm$0.09 & 11.58$\pm$0.02&6.57$\pm$0.002&0.01$\pm$0.01&0.0002(1) & 1.05(424) \\
\rowcolor[gray]{.9}Ns3&
2.11$\pm$0.04&  1.06$\pm$0.01 & 14.0$\pm$0.3 &0.12$\pm$0.04  &11.04$\pm$0.02& 6.78$\pm$0.04&0.01$\pm$0.01&0.0009(1) & 1.14(424) \\
\rowcolor[gray]{.9}Ns4&
 1.94$\pm$0.04&  1.16$\pm$0.03&  16.0$\pm$0.2& -0.38$\pm$0.06 & 12.62$\pm$0.02& 6.71$\pm$0.03&0.03$\pm$0.01&0.0010(3) & 1.09(424)\\
\rowcolor[gray]{.9}Ns5&
2.10$\pm$0.06&  1.07$\pm$0.04&  15.0$\pm$0.3& -0.38$\pm$0.01 & 11.09$\pm$0.01& 6.52$\pm$0.08&0.02$\pm$0.01&0.0007(1) & 1.05(424)\\
\rowcolor[gray]{.9}Ns6&
1.97$\pm$0.06&  1.21$\pm$0.01&  12.0$\pm$0.1& -0.42$\pm$0.01 & 12.42$\pm$0.02& 6.95$\pm$0.07&0.01$\pm$0.01&0.001(1) & 1.13(424) \\
\rowcolor[gray]{.9}Ns7&
 1.95$\pm$0.09&  1.03$\pm$0.02&  18.0$\pm$0.4&  1.12$\pm$0.01 &13.01$\pm$0.12&  6.64$\pm$0.02&0.04$\pm$0.02&0.0034(1)&  1.07(424)\\
\rowcolor[gray]{.9}Ns8&
 1.71$\pm$0.06&  1.02$\pm$0.02&  15.0$\pm$0.2& 2.00$^{\dagger\dagger\dagger}$ &14.21$\pm$0.12&  6.52$\pm$0.08&0.01$\pm$0.01&0.0036(1)&  1.08(549)\\
\rowcolor[gray]{.9}Ns9&
 1.98$\pm$0.06&  1.10$\pm$0.01&  18.0$\pm$0.4& -0.72$\pm$0.01 & 11.64$\pm$0.02& 6.7$\pm$0.1&0.05$\pm$0.03&0.0004(2)&  0.97(424) \\
\rowcolor[gray]{.9}Ns10&
 2.20$\pm$0.09&  1.02$\pm$0.03&  17.0$\pm$0.1& -0.71$\pm$0.01 & 11.89$\pm$0.01&6.7$\pm$0.2&0.03$\pm$0.01&0.0002(2)&  0.97(424) \\
\rowcolor[gray]{.9}Ns11&
 2.20$\pm$0.07&  0.97$\pm$0.01&  18.0$\pm$0.5& -0.41$\pm$0.01 & 11.44$\pm$0.01&6.71$\pm$0.08&0.04$\pm$0.02&0.0002(2)&  1.06(424) \\
\rowcolor[gray]{.9}Ns12&
 2.14$\pm$0.09&  0.97$\pm$0.02&  16.0$\pm$0.2& -0.59$\pm$0.01 & 11.49$\pm$0.01&6.65$\pm$0.07&0.01$\pm$0.01&0.00025(1)&  1.01(424)\\
\rowcolor[gray]{.9}Ns13&
1.91$\pm$0.08&  1.00$\pm$0.01&  16.9$\pm$0.1& -0.74$\pm$0.04 & 11.68$\pm$0.01&6.68$\pm$0.09&0.01$\pm$0.01&0.00025(2)&  1.00(424)\\
\hline
Nc1& 1.0$\pm$0.4&  0.70$\pm$0.01&  6.2$\pm$0.1& -0.28$\pm0.03$ & 7.68$\pm$0.04&7.1$\pm$0.2&0.01$\pm$0.01&0.00001(1)&  1.14(127)\\
\hline
      \end{tabular}
    \label{tab:fit_table_asca_1957}
\\Parameter errors correspond to 1$\sigma$ confidence level.
$^\dagger$ The spectral model is  
{\it tbabs*(CompTB + Gaussian + Bbody(``HTBB''))};
$^{\dagger\dagger}$ normalization parameters of {\it Bbody} and {\it CompTB} components are in units of 
$L_{36}/d^2_{10}$ $erg/s/kpc^2$, where $L_{36}$ is the 
source luminosity in units of 10$^{36}$ erg/s; $d_{10}$ is the distance to the source in units of 10 kpc; 
$T_{bb}$ of $Bbody$ component is low variable from 4.5 keV to 5.7 keV; 
and {\it Gaussian} component is in units of $10^{-3}\times$ total photons cm$^{-2}$s$^{-1}$ in line; 
$^{\dagger\dagger\dagger}$ when parameter $\log(A)\gg1$, it is fixed to a value 1.0 (see comments in the text); 
$N_H$ varies in the range of (0.95 -- 1.7)$ \times 10^{21}$ cm$^{-2}$ 
 \citep[see also][]{Barillier23}. 
\end{table*}

\begin{table}
 \caption{The best fit spectral parameters from IXPE observation of 4U~1957+115 for the spectro-polarimetric model {\tt tbabs*(polpow*(compTB+Gauss))}.  From left to right are, (1) model components; (2) parameters in components; (3) best fit values for Obs. Ix1. The parameters that are fixed during the fits are denoted with $^{fixed}$.}
   \label{tab:table_IXPE}
 \begin{tabular}{lll}
 \hline\hline  
Components &  Parameter & Value \\
 \hline
{\it polpow} &  $A_{norm}$, \% &2.77 $\pm$ 0.32\\
                  & $A_{index}$ & -0.76 $\pm$ 0.07\\
                  & $psi_{norm}$, deg & -41.72 $\pm$ 0.53\\
                  & $psi_{index}$ & 0$^{fixed}$\\
{\it tbabs}          & $N_H$, 10$^{22}$ cm$^{-2}$ & 0.2$\pm$0.1\\
{\it compTB} &$\alpha$& 1.06$\pm$0.01\\
                   &  $kT_s$, keV & 0.7$\pm$0.01\\
                   &  $kT_e$, keV& 6.0$\pm$0.01\\
                   & $\log{A}$  & -0.28$\pm$ 0.03\\
                   & $N_{com}^{\dagger}, L_{36}/d_{10}^2$ &7.8$\pm$0.1 \\
                   & $E_{gauss}$, keV&  6.9$\pm$0.1\\
                   & $N_{gauss}^{\dagger\dagger}$ & 4.15$\pm$0.08\\
                   & $\chi^2_{red}$ (dof)     &  1.32 (447)\\
      \hline
 \end{tabular}
\\$^{\dagger}$ $L_{36}$ is the source luminosity in units of $10^{36}$ erg/s and $d_{10}$ is the distance to the source in units of 10 kpc.  
$^{\dagger\dagger}$ {\it Gaussian} component is in units of $10^{-3}\times$ total photons cm$^{-2}$s$^{-1}$ in line. 
\end{table}

Different X-ray missions have observed 4U~1957+115 during different spectral states at different times, giving us an idea of what is happening to this source in X-rays.

\subsubsection{Evolution of spectral properties during  different  X-ray state transitions \label{spectral analysis_1957}}

Spectral analysis of 4U~1957+115 using observations by ASCA, {\it Suzaku, RXTE, IXPE, Swift, NICER} and {\it NuSTAR} indicates  that  the source  spectra can be  reproduced by a model with an absorbed  Comptonization component, 
with  an addition of {\it Gaussian} iron line component. In the analyzed observations, 4U~1957+115  exhibits enhanced flare activity (see Figs. \ref{ev_1957} and \ref{fraq_1957}), accompanied by a change of  its spectral shape.


4U~1957+115 smoothly evolves between different spectral states, from the LHS ($1.7\le\Gamma<2$ and $20 \le kT_e \le 50$ keV) at the beginning of the outburst, passing through the IS in the middle of the outburst and reaching the HSS ($2.7 < \Gamma < 2.9$ and $3 \le kT_e \le 15$ keV) at the outburst maximum (Fig. \ref{gam_norm_Te}). However, 
in the LHS, along with the spectra typical for this state
we found a number of spectra that are somewhat different from those and appear as a positive excess of soft X-rays at energies  3--5 keV in the source spectrum. We designated such spectra ``soft-LHS'' and associated this excess with the increased efficiency of the accretion disk radiation.
The source 4U~1957+115 in the HSS, along with the spectra typical for this state ($2.7<\Gamma<2.9$), showed a number of spectra that were somewhat different from those, as a 
positive excess of hard X-rays in the source spectrum at energies 15--50 keV. We designated such spectra ``hard-HSS'' and we formally approximated this hard excess by 
a {\tt blackbody} model  with a relatively high color temperature ($kT_s\sim 4-5$ keV, the so-called high-temperature blackbody (HTBB) component). We associate this excess with the effects of the matter converging onto  a BH (this effect is discussed in more detail in a number of papers \citep{ts09,st10,TS21}, as well as in the Sect. \ref{discussion of spectral analysis}). Examples of the spectra in such states are shown in Figs. \ref{spectrum_ev_1957} and \ref{spectrum_ev_1957_HTBB} -- \ref{NuSTAR-spectra_N8}, as well as described using 4U~1957+115 data by different missions in Sects.~\ref{spectral analysis_rxte} -- \ref{spectral analysis_nustar}.  
In the best-fit spectral analysis, $N_H$ varied weakly in the range $(0.9-1.7)\times 10^{21}$ cm$^{-2}$, which is consistent with the $(0.95-1.7)\times 10^{21}$ cm$^{-2}$ estimates from the joint {\it NICER}/NuSTAR analysis of the 4U~1957+115 spectra \citep{Barillier23} and with the Galactic extinction level $2.1\times 10^{21}$ cm$^{-2}$~\citep{W00}.

\subsubsection{Spectral properties of 4U~1957+115 based on {\it RXTE} observations \label{spectral analysis_rxte}}

The evolution of the spectral parameters of 4U~1957+115  in 3--60 keV range during outbursts according to {\it RXTE} data from 1997 to 2010 is shown in Fig. \ref{fraq_1957}. The peaks of the $F_1$, $F_2$ and $F_3$ outbursts are indicated by arrows in the upper part of this figure and the phases of the outburst decay are marked by vertical blue stripes. It is clearly seen that in the LHS of 4U~1957+115 
the hard emission component with low $\Gamma$ values, a high fraction  of Comptonization $f$ and weak X-ray normalizations $N_{com}$ is observed. 

In particular, in the {\it  soft-LHS}, represented by the {\it RXTE} observation (ID=50128-01-05-05), the spectrum 
 is dominated by a soft excess and the hard emission component (Fig. \ref{spectrum_ev_1957}, panel $a$). This hard component is well reproduced by the Comptonization model with  the following parameters: 
 $\Gamma$=1.75$\pm$0.07, $kT_e$=23.8$\pm$0.03 keV, $N_{com}$=7.67$\pm$0.01 $L_{36}/d^2_{10}$ and $T_s$=0.99$\pm$1 keV (reduced $\chi^2$=0.96 for 54 d.o.f). The data  are shown  by black crosses and the best-fit spectral  model   {\tt tbabs*(CompTB+Gauss)} by red line.  In the bottom panels we show  $\Delta \chi$ versus a photon energy in keV. 

In the  panel $b$ (Fig. \ref{spectrum_ev_1957}) we demonstrate  the source spectrum (ID=70054-01-04-00) in the IS, for which the best-fit model parameters  are $\Gamma$=2.17$\pm$0.3, $N_{com}$=12.66$\pm$0.09 $L_{36}/d^2_{10}$, $kT_e$=5.1$\pm$0.09 keV and $T_s$=1.0$\pm$0.3 keV 
($\chi^2_{red}$=0.98 for 54 d.o.f). 

In the HSS, the spectral parameters of the source change remarkably: soft radiation already dominates, $\Gamma$ increases, the  Comptonization fraction,  $f$ decreases with an increase of  normalization, $N_{com}$. For example, in the panel $c$ (Fig. \ref{spectrum_ev_1957}),  the {\it HSS} spectrum (ID=40044-01-03-03) is presented, where we can see  the dominance of the  soft emission, for which the best-fit model parameters  are $\Gamma$=2.10$\pm$0.3, $N_{com}$=13.7$\pm$0.4 $L_{36}/d^2_{10}$, $kT_e$=12.7$\pm$0.4 keV and $T_s$=0.89$\pm$0.1 keV 
($\chi^2_{red}$=0.99 for 54 d.o.f). 

Finally, the panel $d$ of Fig. \ref{spectrum_ev_1957} ({\it hard-HSS}) demonstrates the {\it HSS} source spectrum (ID=70054-01-01-01) 
for which the best-fit model parameters  are $\Gamma$=2.62$\pm$0.3, $N_{com}$=15.66$\pm$0.09 $L_{36}/d^2_{10}$, $kT_e$=6.25$\pm$0.06 keV and $T_s$=0.78$\pm$0.02 keV 
($\chi^2_{red}$=0.95 for 54 d.o.f). 

We clearly see that the change of the spectral state from the LHS to the HSS in 4U~1957+115 is accompanied by an increase of  normalization, $N_{com}$ and an increase in the photon index $\Gamma$ from 2 to 3 (red points in Figure \ref{gam_norm_Te}). 
 However, with a further increase in  normalization (or the mass accretion rate)  the photon index $\Gamma$ drops  from 3 to 2.4.

 It is important to emphasize  that in Figure \ref{spectrum_ev_1957_HTBB} 
 we show a spectral  evolution from the IS to the HSS  where one can clear see a presence of the so-called high-temperature blackbody (HTBB) component, which appears in the IS when $\Gamma\sim 2$ and disappears in the HSS when $\Gamma\sim 3$. 
Below we will show that the shape of this component can be approximated by a hot black body 
with a color temperature $kT_{HTBB}\sim$ 4 -- 6 keV. 

\subsubsection{Spectral properties based on {\it ASCA} observations \label{spectral analysis_asca}}



The  {\it ASCA} observation {\it A1} were carried out in 1994, when 4U~1957+115 was in the  {\it soft-LHS} state. Best-fit parameters of the spectra  in 0.3--10 keV range using the {\tt tbabs*(CompTB+Gauss+Bbody)} model are $\Gamma=1.73\pm 0.04$, $T_e=18.52\pm 0.02$, $N_{com}=6.69\pm 0.03$, $\chi^2_{red}$=1.07 (277 dof) (yellow 
line in Table~\ref{tab:fit_table_asca_1957}).

\subsubsection{Spectral properties based on {\it Suzaku} observations \label{spectral analysis_suzaku}}

During the {\it Suzaku} observations in May -- November, 2010, 4U~1957+115  was in the HSS state. The best-fit parameters of the spectra  in 0.3--10 keV range using the {\tt tbabs*(CompTB+Gauss+Bbody)} model are given in Table~\ref{tab:fit_table_asca_1957}. 
An example of a typical observed spectrum with the best-fit model is shown in Fig.~\ref{Suzaku_spectra_1957}. 

\subsubsection{Spectral properties of 4U~1957+115 based on {\it NICER} observations \label{spectral analysis_nicer}}

All {\it NICER} observations (May 12 -- 23, 2023) mainly capture the 4U~1957+115 outbursts, when the object was in the IS state (Fig.~\ref{gam_norm_Te}). The best-fit parameters of the spectra  in 0.5--12 keV range using the {\tt tbabs*(CompTB+Gauss+Bbody)} model are given in Table~\ref{tab:fit_table_asca_1957}. 
An example of a typical observed spectrum with the best-fit model is shown by red color in Fig.~\ref{polariz} (right panel). 

\subsubsection{Spectral properties of 4U~1957+115 based on {\it IXPE} observations \label{spectral analysis_ixpe}}
During the {\it IXPE} observations, 4U~1957+115 (May 12, 2023) was in the IS--HSS phase. The best-fit parameters of the spectra  in 2--8 keV range using the {\tt tbabs*(CompTB+Gauss+Bbody)} model are given in Table~\ref{tab:table_IXPE}. Here we also present the results of model-dependent analysis by fitting IXPE Stokes $I$, $Q$ and $U$ spectra in XSPEC in frame of {\tt tbabs*(polpow*(compTB+Gauss))} model. An observed spectrum with the best-fit model is shown by green color in Fig.~\ref{polariz} (right panel). 

The analysis of IXPE data shows that the {\tt polpow} model fits the observations better than, for example, {\tt pollin} or {\tt polconst}. On the one hand, this may indicate that polarization properties  are energy dependent. And this somewhat contradicts the conclusions of ST85 that $P$ does not depend on the photon energy. 
On the other hand, the $P(E)$ estimate for 4U~1957+115 was obtained in the time interval when the source underwent significant changes in the spectral state, MJD 60076--60088 (see Table 2 and Fig.~4 in M24). 
 These changes are well tracked, for example, in terms of the spectral hardness coefficient $HR$, and are presented in Fig.~2 in  (M24): 
0.05--0.06 in NICER [$HR=(4-12keV)/(0.3-4keV)$] and 0.03--0.05 in IXPE [$HR=(5-8keV)/(2-5keV)$] observations. 
As already mentioned, the strict constancy of $P$ with energy, according to ST85, can be ensured if the value of $P(E)$ is determined within one spectral state (e.g., HSS), without admixture of source data in other states (e.g., IS or LHS). As is known, at present the results of IXPE observations in the range of 6--8 keV suffer somewhat from poor statistics, especially for the source 4U~1957+115, which makes the presented dependence of $P(E)$ somewhat uncertain. However, IXPE observations already show that PA is independent of energy (M24,  \cite{Kushwaha23}).



\subsubsection{Spectral properties based on {\it Swift} observations \label{spectral analysis_swift}}

All {\it Swift} observations of 4U~1957+115 (2007 -- 2019) mainly cover the time intervals with the flares, when the object underwent the spectral state transition  from the IS to the HSS (Fig.~\ref{ev_1957}),  
accompanied by a change in the spectral state 
(see Fig.~\ref{4_spectra_1957}). 
The spectral analysis of 4U~1957+115 in 0.3--10 keV range within our Comptonization  model showed that the source smoothly evolves 
from the IS ($2.2 < \Gamma < 2.5$ and $10 \le kT_e \le 15$ keV; e.g., grey $S4$ spectrum in Fig.~\ref{4_spectra_1957}) in the middle of the flare and reaching the HSS ($\Gamma \sim 3$ and $kT_e \sim 7$ keV; e.g.,  brown $S1$ spectrum in Fig.~\ref{4_spectra_1957}) at the flare peak. 



\subsubsection{Spectral properties based on {\it NuSTAR} observations \label{spectral analysis_nustar}}

{\it NuSTAR} observations cover a wide time interval from 2013 to 2023 and detect 4U~1957+115 in different spectral states from the  {\it IS} to the {\it hard-HSS}. In particular, the index $\Gamma$ varies around   3  (see Table~\ref{tab:fit_table_asca_1957}, grey box). The typical spectra of the source in 2--60 keV range is shown  in Figs.~\ref{4_spectra_1957}  ($Ns5$ violet and $Ns8$, khaki) and \ref{NuSTAR-spectra_N8}. 
The disk seed photon temperature $T_s$ varied weakly around 1 keV 
during all {\it NuSTAR} observations. 
The plasma temperature $T_e$ is maintained at a moderate level of $T_e\sim 12-20$ keV, although the illumination conditions of the CC vary greatly, which is traced by variations in the parameter $log(A)$ in a wide range, from --0.77 to 2. Also, in the spectrum of 4U~1957+115, an iron line is detected at energies $E_{gauss}$ from 6.5 keV to 7 keV. The normalization of the {\tt CompTB} component is maintained at a moderate level typical for the IS--HSS phase, $N_{com}\sim (0.011 - 0.014)\times L_{39}/d_{10}^2$ (see Table~\ref{tab:fit_table_asca_1957} and Fig.~\ref{NuSTAR-spectra_N8}), indicating a stable mass accretion rate $\dot M$ during the {\it NuSTAR} observations of 4U~1957+115.

Note, that the available {\it NuSTAR} data allows us to study the broadband spectrum of the source (2 -- 70 keV), 
thus, refine the model for fitting the spectrum (see Fig.~\ref{NuSTAR-spectra_N8} and discussion below in Sect.~\ref{HTBBl}). 
Namely, we investigated the contribution of the HTBB component and its features taking into account its transience. We found that sometimes it reaches 0.0036 $L_{36}/d_{10}^2$, and sometimes it is not detected in the spectrum at all (table~\ref{tab:fit_table_asca_1957}). We fixed HTBB at the detection limit in cases where taking HTBB into account is not required when modeling the source spectrum. This approach is justified for the subsequent adequate comparative analysis of the results of fittings on different dates.

\subsection{Observational evidence of the ``BB-like" (HTBB) component peaked at $\sim$20 keV in the IS spectra \label{HTBBl}}

The adopted spectral model shows  a very good agreement for 47 cases out of 80 spectra of 4U~1957+115 with RXTE used in our analysis. A value of the reduced statistic $\chi^2$ ($\chi^2_{red}=\chi^2/N_{dof}$,  where $N_{dof}$ is the number of degrees of freedom for the fit, is  about  1.0. for 47  observations. But for  33 of the IS  observations, the fit of the data with the {\tt tbabs*(CompTB + Gauss)} model is not so good; $\chi^2_{red}$  reaches 1.4 and even higher. We find that there is a characteristic bump in the data residual compared to the model at about 20 keV, which can be fitted by a BB-like shape of the color temperature at about 4.5 keV. This 
HTBB  component is strong in each of the 33 observations (see Fig.~\ref{spectrum_ev_1957_HTBB}). 

A similar pattern can be demonstrated using NuSTAR observations of 4U~1957+115. For example, when modeling the spectrum $Ns8$ with the {\tt tbabs*(compTB + Gauss)} model, the fit is not as good, with $\chi^2_{red}$ reaching 2.1. This spectrum is shown in the left panel of Figure~\ref{NuSTAR-spectra_N8} with the best-fit model (top) and $\Delta\chi$ (bottom). Again, as in some RXTE observations, the data residuals  are found to exhibit a characteristic positive excess around 20 keV compared to the best-fit model (see the lower left panel of Figure~\ref{NuSTAR-spectra_N8}). This bump is easily fitted to the BBody-like shape with the color temperature $\sim$4.5 keV (see the 
right panel of Figure ~\ref{NuSTAR-spectra_N8} and Table ~\ref{tab:fit_table_asca_1957} for the best-fit parameters). Specifically, the right panel of Figure ~\ref{NuSTAR-spectra_N8} (top) shows the best-fit spectrum and $\Delta\chi$ (bottom) when the bump in the residuals at $\sim$ 20 keV is modeled by the HTBB component with $\chi^2_{red} = 1.08$ for 549 dof, 
thus achieving a good statistical fit to the model.

We present the observational results in Figure \ref{gam_norm_Te} where the photon index, $\Gamma$ plotted versus the $CompTB$ normalization, $N_{com}$ (top panel) and  the electron temperature $T_e$ 
(bottom panel) for 4U~1957+115. 
Here we marked two vertical zones that differ in the normalization $N_{com}$(proportional to the mass accretion rate $\dot M$): I -- 
zone of moderate $\dot M$ (pink stripe) and II -- 
zone of high $\dot M$ (blue stripe). These zones show completely different behavior of $\Gamma-N_{com}$: zone I is characterized by a monotonic increase of $\Gamma$ with $N_{com}$, while zone II is distinguished by a decrease of $\Gamma$ with $N_{com}$. The green horizontal stripe indicates the interval for moderate $\Gamma\sim$ 2.2--2.8, where the appearance of HTBB is expected \citep{TS21}. In 4U~1957+115, HTBBs are well detected in the area of intersection of the green and pink stripes, although in the area of intersection of the green and blue stripes HTBB is almost not detected.

This difference is more clear when looking at the spectra of 4U~1957+115 from these zones in detail (Figure \ref{HTBB_I_II}). In fact, the spectra in zone I (enclosed by the red rectangle in this figure) 
show a well-detected HTBB contribution (HTBB is shaded in yellow), while in zone II (enclosed by the green rectangle) the spectrum is well described by the Comptonized model without any HTBB contribution (in these cases, we set the HTBB normalization to a minimum level). 
The key parameter here is the CC plasma temperature (see Fig.~\ref{4_spectra_1957}). So for zone I the temperature $T_e$ is higher ($\sim$20 keV) than for zone II ($\sim$5 keV).
%
The reason for this behavior is discussed in \cite{LT18} and \cite{TS21} and the spectral shape of this feature is presumably related to the gravitational redshift,  $z$ of the AL 
formed  near a BH horizon, where this feature is formed, and the optical depth $\tau$ of the converging flow \citep{LT18}. We discuss the details of this interpretation in Sect.~\ref{HTBB}.

\subsection{A BH mass estimate}
\label{mass_estimate}

We used a scaling technique to estimate a BH mass $M_{BH}$, previously developed specifically for a BH weighing~ [\cite{TSM25,TS24,TS23,TSC23,STU18,SCT18,STV17}]. 
This scaling technique is widely used for a BH mass estimate.  
It is assumed that under outbursting conditions the characteristics of the X-ray emission from the inner region of the disk are similar for different BHs of different masses 
  \citep[][hereafter  ST09]{ST09}, which is consistent with empirical data \citep{SCT18}.  The disk luminosity [see \cite{SS73}] can be evaluated as:

\begin{equation}
L=\frac{G\dot M M_{BH}}{R_{\ast}}, 
\end{equation}

\noindent where $G$ is the gravitational constant, $\dot M$ is the mas s accretion rate onto the BH, $R_{\ast} = r\times R_S$ is the effective radius at which the main energy release in the disk occurs, $R_S = 2G\times M_{BH}/c^2$ is the Schwarzschild radius. In  the framework of this assumption, the {\it target} BH mass,  $M_t$ can be scaled relative to another one,  {\it reference} BH,  $M_r$:
\begin{equation}
 s_N=\frac{N_r}{N_t} =  \frac{m_r}{m_t} \frac{d_t^2}{d_r^2}{f_G},
\label{mass}
\end{equation}
where $N_r$ and $N_t$ are normalizations of the spectra, $m_t=M_t/M_{\odot}$ and  $m_r=M_r/M_{\odot}$ are the dimensionless  BH masses with respect to a solar mass, and $d_t$ and $d_r$  are distances to  the {\it target} and {\it reference} sources, correspondingly.  A   geometrical factor, $f_G=\cos i_r/\cos i_t$, where $ i_r$ and $ i_t$ are the disk inclinations for   the {\it reference}  and {\it target} sources, respectively 
(ST09).


We  apply  the $\Gamma-N$ correlation to estimate the mass of BHs (for details see ST09). 
This method ultimately ({\it i}) identifies a pair of BHs for which $\Gamma$ correlates with an {increasing} normalization of $N$ (which is proportional to a mass accretion rate $\dot M$ and a BH mass $M_{BH}$, see Eq.~(7) in ST09)
and for which the saturation levels $\Gamma_{sat},$ are the same and ({\it ii}) calculates the scaling factor $s_{N}$, which allows us to determine a black hole mass of the target object. 

For appropriate scaling, we need to select X-ray sources (reference sources), which also show the effect of  the index saturation, namely at the same $\Gamma$ level as in 4U~1957+115 (target source). For these reference sources, a BH mass, inclination, and distance must be well known. We found that 
GRS~1915+105, 4U~1630--47 and H~1743--322 can be used as the reference sources because these sources met all aforementioned  requirements to estimate a BH mass of the target source (see  items (i) and (ii) above). 

\begin{figure*}
\includegraphics[scale=0.95,angle=0]{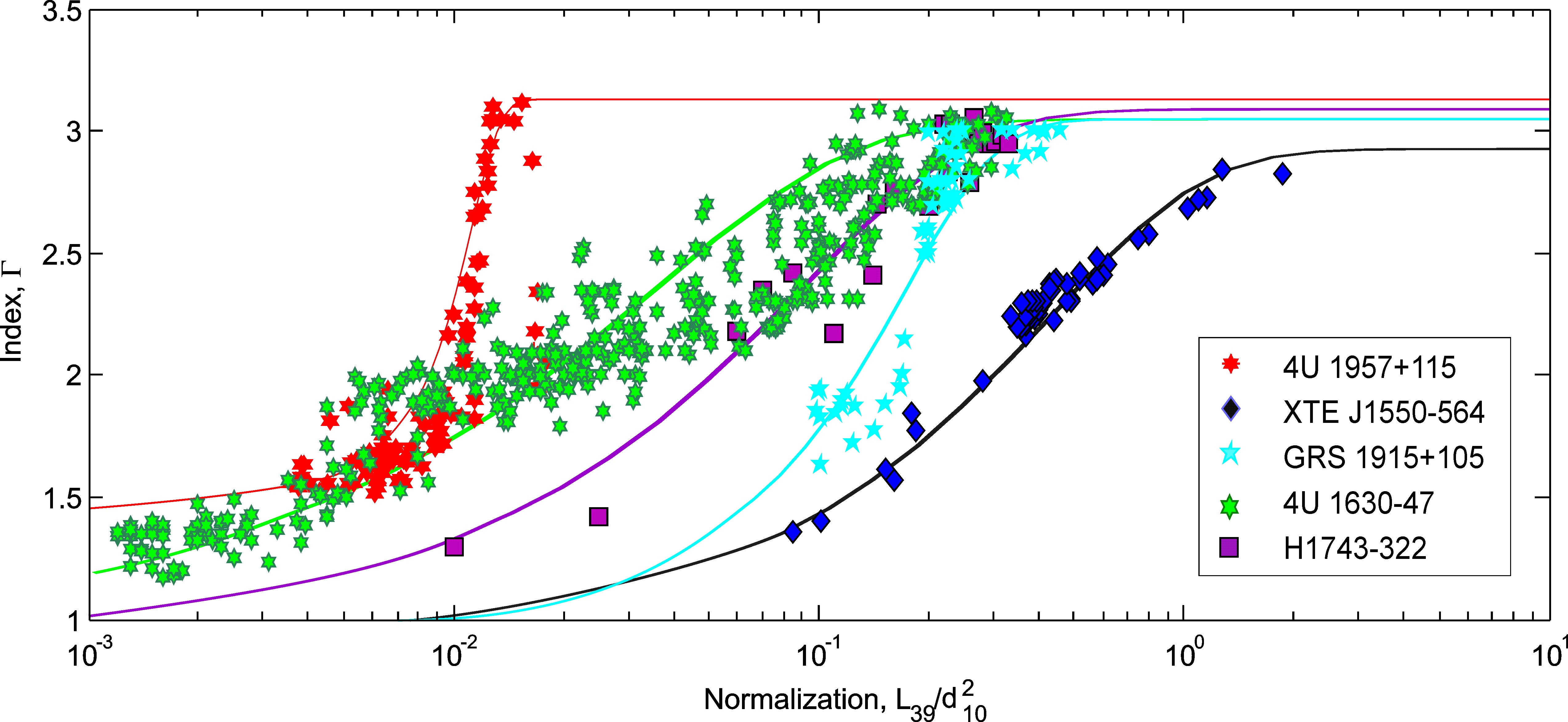}
\caption{
Scaling of photon index $\Gamma$ for 4U~1957+115 (with red line for target source) with 
GRS~1915+105 (bright blue line), 4U~1630--47 (green line) and H~1743--322 (violet line) as reference sources).
}
\label{scaling_1957}
\end{figure*}

In Figure \ref{scaling_1957} we demonstrate how the photon index $\Gamma$ evolves with normalization $N$ (proportional to the mass accretion rate $\dot M$) in 4U~1957+115 ({\it target} source) and 
GRS~1915+105, 4U~1630--47 and H~1743--322 ({\it reference} sources),  where  $N$  is presented in units of $L_{39}/d^2_{10}$ ($L_{39}$ is the source luminosity in units of $10^{39}$ erg/s and $d_{10}$ is the distance to the source in units of 10 kpc). As we can see from this Figure 
that these sources have almost the same index saturation level $\Gamma$. 
We estimated a BH mass for 4U~1957+115 using the scaling approach  (see e.g.,  ST09). In Figure~\ref{scaling_1957} we illustrate  how the scaling method works shifting one correlation versus another.     From these correlations we could estimate $N_t$, $N_r$ for 4U~1957+115 and   for the reference sources (see Table~\ref{tab:par_scal_19577}). A value of $N_t=1.02\times10^{-2} ~L_{39}/d^2_{10}$,  $N_r$ 
 is determined in the beginning of the $\Gamma$-saturation  part (see Fig.~\ref{scaling_1957}, ST07,  ST09, \cite{STS14,TS16,tsei16b,ts09}).

%
%

A value of   $f_G=\cos {i_r}/\cos{i_t} $,  using  the {\it target} and {\it reference} sources  can be obtained,  using an inclination for 4U~1957+115 $i_t=72^{\circ}$  and $i_r$   (see Table \ref{tab:par_scal_19577}).    Thus,  we obtain  an  estimated target BH  mass,  $m_t$   (in 4U~1957+115):
\begin{equation}
m_t= f_G\frac{m_r}{s_N} \frac{d_t^2}{d_r^2}, 
\label{mass_target1}
\end{equation}
where we used   a value of $d_t=22$ kpc  (see Table \ref{tab:par_scal_19577}).

\noindent As a result, we find that M$_{1957}\sim$ 4.8$\pm$1.8 M$_{\odot}$ ($M_{1957} = m_t\times M_{\odot}$) assuming $d_{1957}= 22$ kpc for 4U~1957+115. 
 The final error is calculated as the standard deviation relative to the mean: 
$\sigma  (\bar{m_t})= \sigma/\sqrt{n}=0.45/\sqrt{3}\sim0.27$ (see also Table \ref{tab:par_scal_19577}). 


Since the determination of a BH mass in 4U~1957+115 depends on a precise knowledge of the distance we plotted a BH mass versus distance for a wide range of source distances given in GAIA's analysis (from 7 to 80 kpc, see Fig.~\ref{m_D_1957}). According to this plot, only starting from a source distance of 20 kpc  a BH mass  ($m_t$) in 4U~1957+115 is  greater than 3 M$_{\odot}$, i.e. indicates the presence of a black hole in 4U~1957+115. 
 Therefore, following \cite{Nowak08,Nowak12}, we chose the distances $d_t=22$ kpc,  for which the caculations of the BH mass in 4U~1957+115 are presented in Table~\ref{tab:par_scal_19577}.

On the other hand, we found a saturation phase of the index $\Gamma$ in 4U~1957+115 (Fig. 6), which, according to the spectral signatures of the BH \citep{tz98,ts09,TS21}, immediately indicates the presence of a BH in 4U~1957+115. Then, taking into account the graph in Fig. 8, we can impose a lower limit on the distance to the source: $d_t > 20$ kpc, which is consistent with the results of \cite{Nowak08,Nowak12}.

%
%
\begin{table*}
 \caption{BH mass scaling for 4U~1957+115}
 \label{tab:par_scal_19577}
 \centering 
 \begin{tabular}{lllrrll}
  \hline\hline                        
Reference sources   & $m_r$,  M$_{\odot}$       & $i_r^{(a)}$, deg  & $N_r$,               $L_{39}/d^2_{10}$      & $d_r^{(b)}$, kpc  \\
   \hline
GRS 1915+105 $^{(2)}$ & $12.4\pm 2$ & $60\pm5$ & 0.2$\pm$0.1 & $9.4\pm0.8 $\\
4U~1630--47$^{(3)}$ & 10.0$\pm 0.1$& 70& 0.12$\pm$0.04 & $10\pm 1$\\
H~1743--322$^{(4)}$  &  13.3$\pm3.2$ & 70 & 0.19$\pm$0.02 &  $9.1\pm 1.5$\\   
 \hline\hline                        
Target source   & $m_{t}$, M$_{\odot}$ & $i_t^{(a)}$, deg & $d_t^{(b)}$, kpc &    \\
  \hline
4U~1957+115  & $\sim5.6\times(1\pm 0.45)$ &  72 &  22$^{(5)}$    &    that  using  GRS~1915+105  as a ref. source\\
4U~1957+115  & $\sim4.4\times(1\pm 0.45)$ &  72 &  22$^{(5)}$    &    that  using  4U~1630--47  as a ref. source\\
4U~1957+115  & $\sim4.4\times(1\pm 0.45)$ &  72 &  22$^{(5)}$    &    that  using  H~1743--322  as a ref. source\\
4U~1957+115  & Final  estimate                                      &  72 &  22$^{(5)}$   &   as a standard deviation for a mean: \\
           &  $\sim4.80\times(1\pm 0.39)$ &      &      &   $\sqrt{0.45/3}=0.39$\\
  \hline
 \end{tabular}
\\
(a)  System inclination in the literature and  
(b) source distance found in the literature. 
(1) \cite{Orosz01,Sanchez-Fernandez99,Sobczak99}; 
(2) \cite{Fender99,Greiner01}; 
(3) \cite{STS14};
(4)  \cite{McClintock07}; and 
(5) \cite{Nowak08,Nowak12}.
 \end{table*}


\section{Discussion of the spectral analysis \label{discussion of spectral analysis}} 

\subsection{The index--$\dot M$ correlations: index saturation and decrease at large $\dot M$}

The source, 4U~1957+115 shows X-ray variability on  time scales from weeks to months, being almost constant  at a high flux level from 0.1 to 1$\times 10^{-8}$ erg/s/cm$^2$ in  the  3--10 keV range. 
Based on long-term observations of 4U~1957+115 by the {\it NuSTAR}, {\it RXTE}, {\it Suzaku} and ASCA observatories, we have found two fundamentally different types of outbursts: soft and hard, which belong to different phases (I and II) of the $\Gamma-N_{com}$ correlation.

We have classified four spectral states in this source: the softLHS, IS, HSS and hardHSS (see Fig. \ref{spectrum_ev_1957}).
 It is shown that the spectra of the X-ray source are well described by the Comptonization model with the photon index $\Gamma$ varying from 1.5 to 3. A monotonic increase in $\Gamma$ with increasing accretion rate $\dot M$ and saturation of $\Gamma$ at the level of $\Gamma=3$ at high values of $\dot M$ during outbursts are detected. We attributed the difference in the types of outbursts to  different phases in terms of the $\Gamma-N$ correlation. In addition, we discovered a unique phase of decreasing $\Gamma$ from 3 to 2.4 at very high accretion rates. Such  behavior is typical for a number of other X-ray binary systems with BHs (e.g., GRS~1915+105), based on which a conclusion is also made about the presence of a BH in the source. In addition, we estimated a BH mass in 4U~1957+115 by the scaling method $M_{1957}=4.8 \pm 1.8 M_{\odot}$ assuming the source distance of 20 kpc, with 
H~1743--322, 4U~1630--47 and GRS~1915+105 as reference sources. 

Finally, a significant transient feature was found in the source spectrum at energies of 10--20 keV, which we attributed to a gravitationally redshifted AL 
and approximated by the blackbody  with a temperature of 4--5 keV. This feature is observed in the spectra of source with the photon index $\Gamma\sim 2.4$ and is presumably formed in a layer near the event horizon  in 4U~1957+115. Let us discuss this interpretation in more detail below.

\subsection{ The HTBB in different spectral states\label{HTBB}}

In Figure \ref{gam_norm_Te} we demonstrate  the photon index, $\Gamma$, plotted against the $CompTB$ normalization $N_{com}$ (upper panel) and the electron temperature $T_e$ (in keV, lower panel) for 4U~1957+115. Here we also highlight two zones (I and II) for $\Gamma\sim$ 2.2--2.8, where the HTBB is expected to appear (green band for $\Gamma$). The decay phase of the photon index $\Gamma$ at large normalizations ($N_{com}\ge15\times L_{36}/d^2_{10}$) is indicated by the vertical blue band (zone II). 

The pink vertical band (zone I) marks the normalization interval corresponding to the detection of the HTBB at intermediate accretion rates $\dot M$ ($11\times L_{36}/d^2_{10}\ge N_{com}\ge 13\times L_{36}/d^2_{10}$). It is interesting to note that similar HTBB humps was previously detected in other BH sources {Cyg~X--1 \citep{Tomsick18, TS21}, GX~339--4 \citep{TS21}, Cyg~X--3 \citep{Koljonen13,Shrader10}, GRS~1915+105, SS~433 \citep{st10,TS21}, GS~2000+25, GS~1124--68 and XTE~J1550--564 \citep{Zycki01}  and V4641~Sgr \citep{TS21}) in the X-ray spectra in its IS states (for the photon index $\Gamma >$ 1.9) under the conditions of zone I \citep{TS21,LTS22}. 
However, the question arises as to why HTBB hump is minimal or absent under the conditions of zone II at the same photon indices (Figs.~\ref{gam_norm_Te} and \ref{HTBB_I_II}). 
These results raise a valid question about the nature of the HTBB component as an intrinsic property of a BH. 

 We found a difference in the spectra  from the zone I (enclosed 
in red) and zone II (enclosed 
in green in Figure~\ref{HTBB_I_II}) for  the same $\Gamma =2.2-2.8$, but for different normalizations $N_{com}$. Indeed,  the HTBB is easier to detect in zone I, while it is barely seen   or absent at  in zone II. 
It is obvious that in zone I the HTBB contribution is well detected, while in zone II the spectrum is well described by the Comptonized model with  a minimal HTBB contribution. We demonstrate this difference using the zone I spectra 50128-01-03-00 (left) and 50128-01-02-11 (right) and the zone II spectra 40044-02-01-00 (left) and 40044-02-01-01 (right)). The data are represented by crosses, and the best spectral model {\tt tbabs*(CompTB+BB)} is shown by  the red line. The yellow shaded areas show an  evolution of the HTBB component during the source transitions between the LHS and the  HSS. 

Among the possible causes of the formation of the ``humps'' are the following: 1) Compton reflection, 2) photoelectric absorption of photons below 10 keV in a cold medium (disk), 
and 3) a gravitationally redshifted AL. 
Let us consider their possible contribution to the radiation separately.

1) The suggestion that the HTBB hump observed as a positive 20 keV excess in the spectrum is a signature of the Compton reflection (see e.g. \cite{Basko74,ST80,ct95,Magdziarz+Zdziarski95} faces difficulties given that the hard power-law tails of these spectra are too steep to form a Compton hump. Indeed, \cite{ST80} and later \cite{LT07} demonstrated that a Compton hump resulting from photon accumulation due to hard photon scattering in a cold medium (e.g. a disk) cannot be obtained if the photon index  
is $\Gamma > 2$.

2) In principle, the HTBB hump could be a result of photoelectric absorption of photons below 10 keV in the cold medium (disk) given the photoelectric absorption cross-section $\sigma_{ph} \sim (7.8 keV/E)^3\sigma_T$, where $E$ is the photon energy and $\sigma_T$ is the Thomson cross section (e.g. \cite{ct95}).  \cite{Laming+Titarchuk04} 
showed  analytically  
that this reflection hump is not formed if  the photon index $\Gamma>2$.

3) An assumption that the hump of the BH  spectrum is caused by a gravitationally redshifted AL 
is based on the  fundamental theoretical conclusions of the General Relativity. It is known that near the event horizon of a BH, conditions arise for the formation of pairs of electrons and positrons with their subsequent annihilation, accompanied by the emission of photons with an energy of 511 keV. This effect is associated with  a formation of the 
AL. In this case, the gravitational field of a BH is so strong that the initial energy of photons, as they move away from the event horizon of a BH, decreases to $\sim$20 keV. The line itself is blurred due to the non-zero thickness of the formation layer (for example, $\sim$100 m for a BH of a mass of 10 M$_{\odot}$) and turns into a wide ``hump'' in the spectrum at 15--40 keV.  In fact, \cite{LT18} showed that photon-photon interactions of efficiently energy-scattered photons lead to powerful pair production near the BH horizon. Indeed, most of the inwardly scattered photons are deflected outward by relativistic free-falling electrons (light aberration effect; see, e.g., \cite{Rybicki+Lightman79} and Appendix A in \citep{ts09}). These outwardly deflected from a BH scattered photons with energy $E_{up}$ interact with the incoming flux photons with energy $E_{in}$, and this interaction eventually leads to pair production if the condition $E_{up}E_{in} \ge (m_ec^2)^2$ is satisfied.  
In this case, the converging flow  moving with the free-fall velocity $v$ into  a BH with the Lorentz factor $\gamma = 1 + z = 1/\sqrt{1 - (v/c)^2}\gg 1$ should reach regions very close to the BH event horizon, $R = R_S + \Delta R \approx R_S(1 + \gamma^{-2})$ \citep{ts09}, where $\Delta R = R - R_S$ is the radial distance to the BH horizon. Thus, the created positrons actively interact there with the electrons of the accreting flow, and therefore the photons of the AL 
are created and distributed in a relatively narrow shell near the BH horizon. In particular, $\Delta R \le 3 \times 10^4(10/\gamma)^2 ({\cal M}_{BH}/1 M_{\odot}$) cm, where ${\cal M}_{BH}$ is the BH mass expressed in M$_{\odot}$ \citep{ts09}. Then the proper energy (in the convergent flow to the BH) of the photons of the AL 
($E_{511}$) should be visible to the  Earth observer  
at energies with a redshift of up to
\begin{equation}
E^{obs}_{511} = (1 - R_S/R)^{1/2}E_{511} \approx (\Delta R/R_S)^{1/2} E_{511} \approx E_{511}/\gamma.
\end{equation}

In result, a significant fraction of these AL photons, strongly gravitationally redshifted, can be directly visible to an observer on Earth as a hump located at $\sim$20 keV. 
For example, for a BH with a mass of 10 M$_{\odot}$, $\Delta R \sim 3 \times 10^4 \times (10/\gamma)^2 \times 10\sim 
100~m$, hence $\gamma = 17$.  
\cite{LT07} 
performed an extensive Monte Carlo simulation of the formation of the X-ray spectrum in a flow converging onto  a BH, taking into account photon-electron, photon-photon and pair-electron interactions, which confirms the feasibility of the proposed interpretation in case (3). Indeed, in some cases, the spectra of accreting BHs formed in IS and HSS contain a red-shifted AL located at $\sim$20 keV, which can be approximated using the HTBB component.

Interestingly, the NuSTAR spectra of 4U~1957+115 (e.g., 
$Ns8$) have also been successfully modeled previously, e.g. by \cite{Draghis23,Barillier23} 
using relativistic X-ray 
reflection models ({\tt relxill, rellxillCp}). 
We presented here a slightly alternative approach to fitting these data taking into account various physical processes, for which we proposed {\tt compTB} with additional HTBB. We argued that fits using this model in a wide energy range also lead to good results (right panel of Fig.~\ref{NuSTAR-spectra_N8}), and also provide motivation for a self-consistent account of these physical processes occurring in 4U~1957+115 during spectral transitions in X-ray outbursts. 
It is well known that if the photon index $\Gamma$ is higher than 2, there is no reflection hump \citep[see][and also ST80]{LT07}. 
For example, in Fig.~\ref{NuSTAR-spectra_N8} 
the photon index of the source spectrum is significantly higher than 2, so the observed hump at $\sim$20 keV cannot be associated with reflection. Therefore,  taking into account the arguments given in \citep{LT18}, we should accept this hump as due to the formation of an AL 
near the BH horizon.

\subsection{Interpretation of the  index saturation and decrease at large $\dot M$\label{HTBB_interprate}}

In our study for X-ray observations of 4U~1957+115, we found the correspondence of its X-ray variability with a standard pattern $\Gamma-N_{com}$, typical of accreting BHXRBs during outbursts. 
Namely, the photon index, $\Gamma$ increases with normalization $N_{com}$ (which is proportional to $\dot M$)  for moderate $\dot M$ and saturates at high $\dot M$ (zone I in Figs.~\ref{gam_norm_Te} and \ref{HTBB_I_II} (in left red box)). 
However, after reaching the index saturation phase ($\Gamma\sim 3$ ) and further increase in normalization $N_{com}$, $\Gamma$ drops to values of about $\Gamma=2$ at even higher $\dot M$ (zone II in Figs.~\ref{gam_norm_Te} and \ref{HTBB_I_II} (in right green box)). In fact, we have never seen such a behavior of $\Gamma-\dot M$ before (see Fig.~\ref{gam_norm_Te}). 

This surprising behavior of $\Gamma$ versus $\dot M$ can be explained by a simple scenario where the innermost accretion region (CC) 
consists of two layers moving towards the BH, but with different accretion rates. This behavior was predicted by hydrodynamic modeling and its implications for the X-ray observable properties of BHs \citep{ct95}. They found that one of these flows  moved  with a very high velocity  directly  to  a BH  not forming a shock  while another one proceed through the disk   and finally forming   the shock  located within 5$-$10 Schwarzschild  radii from the central object 
\citep[see Fig.~1 in][]{ct95}.  Because the nature of the flow, there are not enough soft disk photons to illuminate the hot CC. 
In addition,  the Thomson optical depth of the converging flow is quite high,  more than $\tau=3$ and we do not observe the redshifted AL 
formed near a BH horizon ($\sim 200$ meters from it) at very high accretion rates $\dot M$ \citep{LT11}. 

%
%
 \begin{figure}
\centering

\includegraphics[scale=0.65,angle=0]{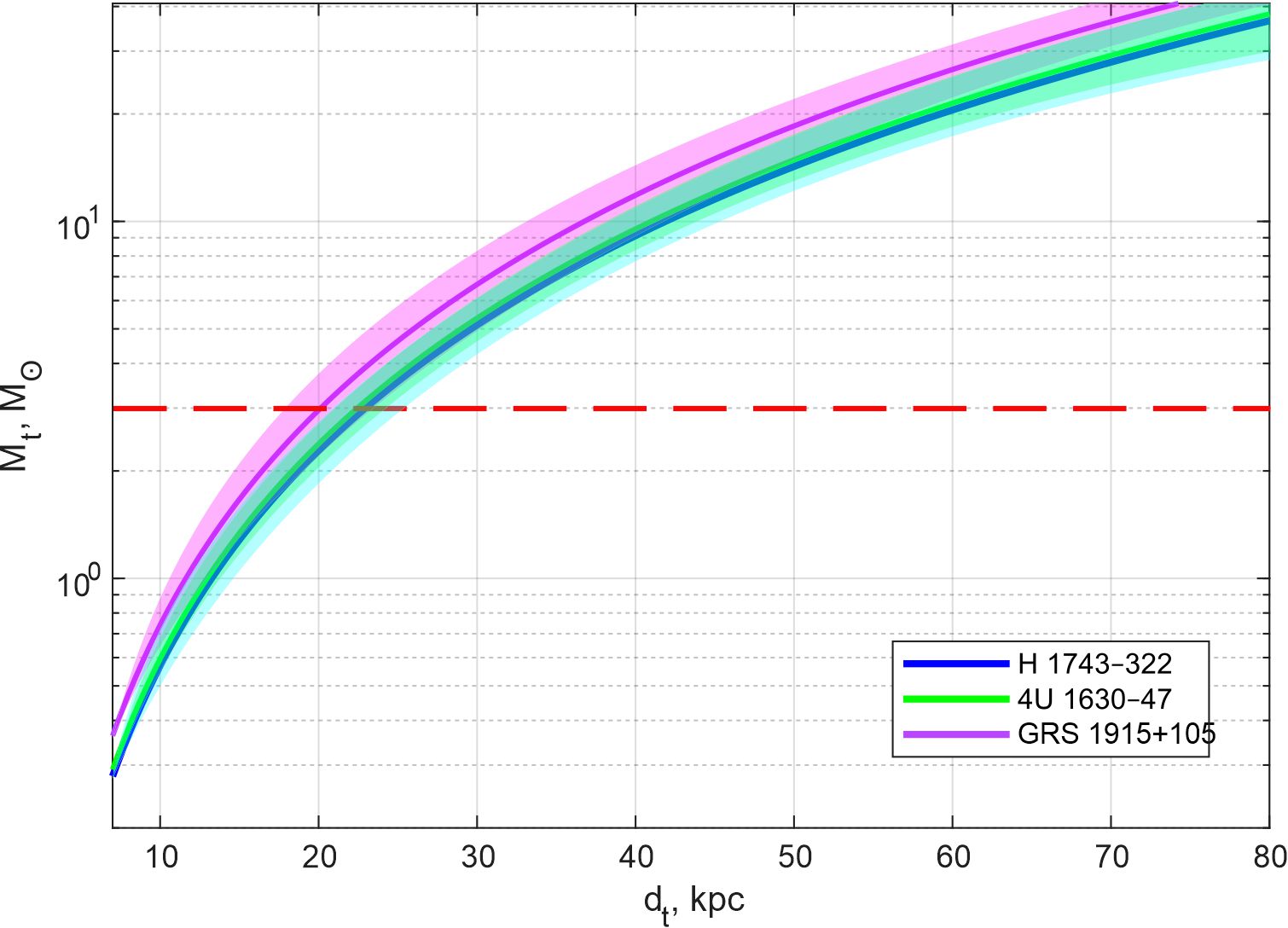}
\caption{
A BH mass $m_t$ versus a distance for a wide range of target source distances given by GAIA's analysis for 4U~1957+115 (red) using 
GRS~1915+105 (violet), 4U~1630--47 (blue) and H~1743--322 (green) as the reference sources.  The corresponding color bands indicate the areas of 1-$\sigma$ estimation errors for each source. Obviously, a BH mass in 4U~1957+115 is more than 5 M$_{\odot}$ (indicated by a red horizontal line) for a source distance greater than 20 kpc.
}
\label{m_D_1957}
\end{figure}

\section{Conclusions \label{summary}} 

A  theoretical   idea of  the LP 
formation in a BH  relies  on  the ST85 study.  The X-ray LP radiation is a result of the multiple  up-scattering   of the initially  X-ray soft photons environment  of  the hot  CC. 
Multiple scattering  of these soft photons leads  to the  formation of the emergent  Comptonization spectrum.  In addition this specific spectrum is linearly  polarized if  the CC  has  a flat geometry.  The polarization degree  of these multiple   up-scattered photons $P$  should be  independent of  the photon energy. 

ST85 calculated the  radiative transfer of the polarized radiation using the iteration method.  A number of  these iterations should be  much more than average number of scatterings in a given flat CC.  As a result, we use  the $P-\mu(i)$ diagram, 
where $i$ is  the source inclination. 
This plot  for  $P(i)$ in $\%$ is calculated  for  different  Thomson optical  depths  $\tau_0$  of the flat CC  from 0.1 to 10    
(see also Figs. 5 and 8 in ST85).

Recent IXPE observations revealed the polarization of X-ray radiation of quite a lot of BHXRBs. For the BH source 4U~1957+115, M24 detected a polarization of radiation in HSS at the level of $P\sim 1.9\%$. Given this new information on $P$ and using the $P (i)$ diagram 
for a given source  we estimate the CC optical depth $\tau_0$. Then, in combination with the results of spectral analysis, we obtain  the photon index $\Gamma$,  the plasma temperature $kT_e$ without any free parameters.


Based on long-term observations of 
4U~1957+115 using  the {\it NuSTAR, RXTE}, {\it Suzaku} and {\it ASCA} observations,
we studied its X-ray variability and detected changes in  the spectral states during source outbursts. 
In particular, the existence of two fundamentally different types of outbursts was confirmed: soft and hard, which belong to different phases (I and II) of $\Gamma-N_{com}$ correlation. 
It was shown that the X-ray spectra of 4U~1957+115 
are well described by the Comptonization model with the photon index $\Gamma$ varying from 1.5 to 3. A monotonic increase in $\Gamma$ with 
mass accretion rate $\dot M$ and saturation of $\Gamma$ at a level of $\Gamma=3$ at high values of $\dot M$ during X-ray source outbursts 
were detected. 
We demonstrated that this behavior of 4U~1957+115, based on a large set of its X-ray observations, is typical of the majority of BHXRBs.
 
We estimated a BH mass in 4U~1957 using  a scaling method $M_{BH}=4.8 \pm 1.8 M_{\odot}$ assuming the source distance about 20 kpc, based on 
 H~1743--322, 4U~1630--47 and GRS~1915+105 as reference sources.   Moreover, a significant transient feature was found in the source spectrum at energies of 10--20 keV, which we attributed to the gravitationally redshifted AL 
and approximated by the blackbody law with a temperature of 4--5 keV. This feature is observed in the source spectra  with the photon index $\Gamma\sim 2.4$ and is presumably formed in a layer near the  BH 
horizon in 4U~1957+115.  We  also found a unique phase of  the $\Gamma$ decrease from 3 to 2.4 at very high accretion rates. 

\begin{acknowledgements}
We appreciate valuable remarks  by Chris Schrader on the paper.    We also recognize the deep understanding of the content of the manuscript  by the referee. 
\facilities{
 MAXI, IXPE, NuSTAR, NICER, RXTE, Swift(XRT), Suzaku, ASCA.}
\software{
HEAsoft 
(HEASARC 2014), XSPEC 
\citep{Arnaud99}, CompTB \citep{Farinelli08}, 
FTOOLS \citep{Blackburn99}, 
ASCASCREEN \citep{Tanaka94}, 
{\it Suzaku} pipeline 
\citep{Mitsuda07},
NuSTARDAS 
\citep{Harrison13}, 
IXPEOBSSIM \citep{Baldini22}.
}
\end{acknowledgements}






\appendix

\section{Data Description}

\subsection{ \it Suzaku data 
\label{suzaku data}}

{\it  Suzaku} observed 4U~1957+115 on 2010 May 4 -- 17 and November 1. Table~\ref{tab:table_Suzaku+ASCA_SAX} summarizes the start and  exposure times, and the MJD interval  for each of these observations. 
One can see a description of the {\it Suzaku} experiment in \cite{Mitsuda07}. 
For observation obtained by a focal X-ray CCD camera (XIS, X-ray Imaging Spectrometer, \cite{Koyama07}), which is sensitive over  the 0.3--12~keV range, we used  software of  the {\it Suzaku} data processing {\tt pipeline} (ver. 2.2.11.22).  We carried out the data reduction and analysis following the standard procedure using the 
{\tt HEASOFT software package} 
and following  the {\it Suzaku} Data Reduction Guide\footnote{http://heasarc.gsfc.nasa.gov/docs/suzaku/analysis/}. 
The spectra of the source were extracted  in 0.3--10 keV range using spatial regions within the 4 ${'}$-radius circle 
centered on the source nominal position (Table~\ref{tab:parameters_binaries}),  while a background was extracted from source-free regions 
for each XIS module separately.
  The spectrum data were re-binned to provide at least 20 counts per spectral bin to validate the use of the $\chi^2$-statistic. We carried out  spectral fitting  applying XSPEC v12.10.1.  The energy ranges around of 1.75 and 2.23 keV are not  used for spectral fitting because of the known artificial structures in the XIS spectra around the Si and Au edges.  Therefore, for spectral fits we have chosen  the 0.3 -- 10 keV  range  for the XISs (excluding 1.75 and 2.23 keV points).  In Fig. \ref{ev_1957} we show  a  light evolution  curve of 4U~1957+115 whereas green arrows indicate the {\it Suzaki}  observations, listed in Table \ref{tab:table_Suzaku+ASCA_SAX}.

\subsection{{\it ASCA data} \label{asca data}}
{\it ASCA} observed 4U~1957+115 on 1994 October 31 -- November 1 with total exposure $\sim$35 ks (see Table~\ref{tab:table_Suzaku+ASCA_SAX}, which  summarized the start time, exposure time, and the MJD interval). One can see a description of the {\it ASCA}  data  by \cite{Tanaka94}.  The solid imaging spectrometers (SIS)  operated in Faint CCD-2 mode. The {\it ASCA} data were screened using the standard processing software (ASCASCREEN and FTOOLS) 
and the standard screening criteria. The spectrum for the source were extracted in 0.3--10 keV range using spatial regions with a diameter of 4${\tt '}$ (for SISs) and 6${\tt '}$ (for GISs) centered on the nominal position of the source,
while background was extracted from source-free regions of comparable size away from the source. The spectrum data were re-binned to provide at least 20 counts per spectral bin to validate the use of the $\chi^2$-statistic. The SIS and GIS data were fitted using {\tt XSPEC} in the energy ranges of 0.6 -- 10 keV and 0.8 -- 10 keV, where the spectral responses are well-known.  

\subsection{\it RXTE data \label{rxte data}}

We have also  analyzed the  available data of 4U~1957+115 obtained with {\it RXTE}~\citep{bradt93} 
(see Table~\ref{tab:par_RXTE_1636}). Standard tasks of the LHEASOFT/FTOOLS 
 software package were utilized for data processing using methods recommended by {\it RXTE} Guest Observer Facility according to the  {\it RXTE} Cook Book\footnote{http://heasarc.gsfc.nasa.gov/docs/xte/recipes/cook\_book.html}. 
For spectral analysis, we used  data from the Proportional Counter Array (PCA) and High-Energy X-Ray Timing Experiment (HEXTE) detectors. {\it RXTE}/PCA spectra ({\it Standard 2} mode data, 3 -- 50~keV energy range) have been extracted and analyzed using the PCA response calibration (ftool pcarmf v11.1). The relevant deadtime corrections to energy spectra have been applied. In turn, HEXTE data were used for spectral analysis only in the 20--50 keV energy range to exclude channels with the highest uncertainty.
We subtracted background corrected in off-source observations.  In Figure  \ref{fraq_1957}  we present  the {\it RXTE} flux in the 3--50 keV ranges and the observational best-fit spectral characteristics. Here we see that the source experienced three flares ($F1$ in 1998, $F2$ in 2002 and $F3$ in 2005), marked with arrows at the top. Systematic error of 0.5\% was applied to the analyzed spectrum. 

\subsection{Swift data\label{swift data}}
 Using {\it Swift}/XRT data \citep{Evans07,Evans09} in 0.3--10 keV energy range we studied a total of five observations of 4U~1957+115  during its flaring events from 2007 to 2019 (Table~\ref{tab:table_Suzaku+ASCA_SAX}).  The data used in this paper are public and available through the GSFC public archive. 
Data were processed using the HEASOFT, 
the tool {\tt xrtpipeline} 
and the calibration files (CALDB version 4.1). The ancillary response files were created using {\tt xrtmkarf} v0.6.0 and exposure maps generated by {\tt xrtexpomap} v0.2.7. Source events were accumulated within a circular region with radius of 46{\tt"} centered at the position of 4U~1957+115 (Table \ref{tab:parameters_binaries}). 
The background was estimated in a nearby source-free circular region of 86{\tt"} radius. 
Spectra were re-binned with at least 10 counts in  in each energy bin using the {\tt grppha} task in order to apply $\chi^2$-statistics. 

\subsection{IXPE data\label{ixpe data}}
We have also  analyzed the  available data of 4U~1957+115 obtained with {\it IXPE} \citep{Weisskopf22} at epoch 1, corresponding to the IXPE observation ID=02006601 ($Ix1$ in Fig.~\ref{ev_1957}). 
From cleaned level 2 event data for each gas pixel detector 80{\tt"} and 60{\tt"} apertures were used to extract source data in 2--8 keV range for polarimetric and spectroscopic analysis, respectively, using the XSELECT 
software applying  the effective eventnumber weighting (STOKES = “NEFF”) to produce 
products in Stokes $I$, $Q$, and $U$ parameters. 
A background region was extracted from an annulus with inner and outer radii of 132{\tt"} and 252{\tt"}, centered on the source position (see Table~\ref{tab:parameters_binaries}). 
Data were extracted into ``polarization cube'' structures, allowing ready data slicing by detector, time, energy, etc. 

\subsection{{\it NICER} data\label{nicer data}}
NICER observed the source from 12 May, 2023 (MJD 60076) to 23 May, 2023 (MJD 60087),  i.e. during the entire IXPE campaign (see Table \ref{tab:table_Suzaku+ASCA_SAX} and Figure \ref{ev_1957}). Eleven observations during this interval are analyzed covering the rise-peack-decay phase of the outburst. 
The {\tt NICERDAS} software 
is used along with the latest CALDB to reduce the data from the observations. 
The {\tt nicerl2} task is used to perform standard calibration and screening to generate cleaned event lists. The source and background spectra along with the responses are generated in the 0.5--12 keV energy band using the {\tt nicerl3}-spect task. The spectra are rebinned to have a minimum of 20 counts per energy bin for spectral modelling. 


\subsection{{\it NuSTAR} data\label{nustar data}}
We processed {\it NuSTAR} observations using the {\tt NuSTARDAS} 
and the latest files available in the {\it NuSTAR} Calibration Database  \footnote{http://heasarc.gsfc.nasa.gov/FTP/caldb/data/nustar/fpm/}. We generated clean event files for each observation of 4U~1957+115 using the {\tt nupipeline} task and extracted the source and background spectra {in 2--70 keV range} from circular regions of 60{\tt"} and 90{\tt"} radius, respectively, centered at the source position (Table~\ref{tab:parameters_binaries}). Source spectra 
were extracted using the {\tt nuproduct} task. We binned each spectrum so that there were at least 25 counts per spectral bin. In Figure \ref{ev_1957}  we indicate using  red arrows particular data when 4U~1957+115 was observed by  {\it NuSTAR}.


\end{document}